\documentclass[journal=jpcbfk,manuscript=article]{achemso}

\usepackage{amssymb,}
\usepackage[version=3]{mhchem} % Formula subscripts using \ce{}
\usepackage[T1]{fontenc}       % Use modern font encodings
\usepackage{xspace}
\usepackage{xr}
\usepackage{cleveref}
\newcommand{\hide}[1]{}
\usepackage{color}

\newcommand{\beginsupplement}{%
        \setcounter{table}{0}
        \renewcommand{\thetable}{S\arabic{table}}%
        \setcounter{figure}{0}
        \renewcommand{\thefigure}{S\arabic{figure}}%
     }

\newcommand{\etal}{{\it et~al.}\ }

\newcommand{\concnorm}[1]{\left[ #1 \right]}

\newcommand{\df}[1]{\mathrm{d}#1}
\newcommand{\dv}[2]{\frac{\df#1}{\df#2}}

\let\oexp\exp
\renewcommand{\exp}[1]{\oexp\!\left\{#1\right\}}

\author{Nathan L Clement}
\altaffiliation{Both authors contributed equally to this work}
\author{Muhibur Rasheed}
\altaffiliation{Both authors contributed equally to this work}
\author{Chandrajit L Bajaj}
\email{bajaj@cs.utexas.edu}
\affiliation[University of Texas]
{Department of Computer Science, The University of Texas at Austin, Austin, TX 78712}

\title[Uncertainty Quantified Viral Assembly]{Uncertainty Quantified Computational Analysis of the Energetics of Virus Capsid Assembly}
\begin{document}
\maketitle

\begin{abstract}
Most of the existing research in assembly pathway prediction/analysis of virus capsids makes the simplifying assumption that the configuration of the intermediate states can be extracted directly from the final configuration of the entire capsid. This assumption does not take into account the conformational changes of the constituent proteins as well as minor changes to the binding interfaces that continues throughout the assembly process until stabilization. 
This paper presents a statistical-ensemble based approach which provides sufficient samples of the configurational space for each monomer and the relative local orientation between monomers, to capture the uncertainties in their binding and conformations. Furthermore, instead of using larger capsomers (trimers, pentamers) as building blocks, we allow all possible sub-assemblies to bind in all possible combinations. We represent this assembly graph in two different ways. First, we use the Wilcoxon signed rank measure to compare the distributions of binding free energy computed on the sampled conformations to predict likely pathways. Second, we represent chemical equilibrium aspects of the transitions as a Bayesian Factor graph where both associations and dissociations are modeled based on concentrations and the binding free energies.
%A maximum a posteriori (MAP) analysis of this graphical model produced expected patterns, e.g. dimer production at a fast rate initially and then consumption as other larger sub-assemblies are created, but also
Results from both of these experiments showed significant departure from those one would obtain if only the static configurations of the proteins were considered. Hence, we establish the importance of an uncertainty-aware protocol for pathway analysis, and provide a statistical framework as an important first step towards assembly pathway prediction with high statistical confidence. 
\end{abstract}
\newpage

\section{Introduction}
Viruses are the smallest living organisms on earth, possessing only a minimal genome which can transcribe only a few proteins. Even with such limited resources, viruses exhibit a remarkable ability to not only survive, but parasitically multiply with great efficiency. One phase of their chemical proliferation that is relatively unexplained, is the spontaneous assembly/disassembly of hundreds of (capsid) proteins that come together to form the capsid (shell) that encases the viral nucleic acid genetic material. Researchers continue to analyze this remarkable process from different perspectives, and also aim to use the insights discovered in designing nano-scale cages and shells for drug delivery \cite{LaiStructure,ZochowskaAdenovirus,ChengDrugDelivery,Smith_2013}. In this article, we present a statistical methodology to analyze the viral assembly process from a free energy perspective. We consider in particular the case of the Nudaurelia Capensis virus (PDBID:1OHF\cite{10hf_Helgstrand}). While others have taken a similar perspective before (e.g.\ \cite{hagan06,hagan2014modeling,zlotnick94,Zlotnick_Mukhopadhyay_2011,rapaport1999}), we are the first to consider positional and conformational uncertainties of the protein structure and their propagated influence on the configurational energetics and binding affinity calculations. This methodology then allows us to infer energetically favorable viral capsomer configurations together with improved statistical confidence. 

Research at understanding the assembled arrangement of capsid proteins in a viral capsid builds upon the work of Caspar and Klug (C-K)\cite{caspar62}. C-K characterize the symmetric organization of proteins in a spherical viral capsid, building upon the mathematical foundations of spherical tilings given by Goldberg\cite{Goldberg_1937}. C-K show that the combinatorial arrangement of the capsid proteins can be characterized using simple triangular tiles that cover an icosahedron. The basic concept is to unfold an icosahedron, which has 20 regular triangular faces (and 12 vertices of 5-fold rotational symmetry), onto a regular hexagonal grid (geometric dual to a triangular grid). By controlling the scale of the icosahedral triangles with respect to the dual-grid-triangles, each icosahedral triangle gets covered by $T$ small {\it grid-triangles}, such that $T = h^2 + k^2 +hk$ where $h$ and $k$ are integers. C-K suggested that each such grid-triangle consists of 3 proteins in a cyclic configuration (forming a trimer). Additionally, there are exactly 12 locations where 5 such small triangles meet at a vertex (vertices of the icosahedron), and $10(T-1)$ locations where 6 such triangles meet. If we only consider the viral protein (of the trimer) which is closest to these 5-fold and 6-fold locations, then they can be thought to produce pentamers (pentons) and hexamers (hexons). Essentially, the entire capsid can be thought of as $20T$ trimers; or as 12 pentamers and $10(T-1)$ hexamers. Finally, their idea, called ``quasi-equivalence,'' states that all the viral protein chains that form a capsid are identical and have (quasi-) equivalent interfaces---essentially, all proteins are involved in the same number of interactions at similar binding sites. 

Recent published work by Janner \cite{janner06} and Keef and Twarock \cite{keef09} have shown that other types of aperiodic arrangements involving pentamers or other types of subassemblies are also possible. Other work, for example by Pawley\cite{Pawley_1961}, shows that several other symmetry classes also permit decomposition into symmetric subassemblies, and Rasheed \etal\cite{rasheed2015highly} proved necessary and sufficient conditions for such subassemblies to be possible. Brooks \etal, in a series of papers, have characterized the geometric conditions for symmetric capsids, provided methods to measure how much a specific capsid conforms to the concept of quasi-equivalence \cite{damodaran2002general,Carrillo-Tripp_Brooks_Reddy_2008,mannige08}, and hence how amenable it is to coarse-grained dynamics analysis as described below. Furthermore, they present a simple classification which characterizes variations of hexamers within a capsid \cite{Mannige_Brooks_2010}. 

Many of the researchers working on predicting, analyzing, and/or simulating capsid assembly have taken either a set of trimers, or a set of pentamers+hexamers, as the building blocks of assembly. For instance, Rapaport \etal\cite{rapaport1999} performed a coarse dynamics simulation where trimers were used as building blocks. The shape of the trimers were modeled using a collection of large balls, and the interactions were modeled by proximity of some small balls placed strategically along the binding region. While this is a very simplistic model, it was successfull is showing that such shape and binding site conditions are sufficient to drive self-assembly. In their more recent work (e.g. \cite{Rapaport_2004,Rapaport_2010}), the model was updated to include more complex energetics, and single proteins (monomers) were used as building blocks, instead of trimers.

Hagan \etal\ \cite{hagan06,Elrad_Hagan_2008} applied Brownian dynamics simulation with a simplified force field. They modeled each capsomer (which can also be a monomer) using a single bead model. Based on prior knowledge about the arrangement of such beads on the capsid, they parametrized each bead based on the angles between each pair of their neighbors, and designed a binding affinity function which allowed binding at specific orientations. The objective of the study was to gain insight from exploring the energy landscape of the assembly, and to identify kinetic traps, analyze the rate of assembly etc. This concept is similar to the `local-rules' introduced by Berger \etal\cite{berger94,berger94b}, which has been adopted by other groups\cite{xie2012surveying,schwartz98}, for kinetics and dynamics analysis of capsids. A discussion contrasting the block-like beads used by Rapaport \etal with shape-driven assembly, and the ones used by Hagan \etal with neighborhood-driven assembly can be found in \cite{hagan2014modeling}. Bona and Sitharam also considered a bead-like model \cite{Bona_Sitharam_Vince_2011}; however, they expressed the interaction between beads using geometric stability conditions and predicted likelihood of binding based on the simplicity of solving the resultant geometric constraints system \cite{sitharam2012easal}.

Unlike the dynamics-based analysis techniques described above, Zlotnick applied statistical thermodynamics law of mass action to relate the concentrations of the constituents and the product of a binding with the binding free energy \cite{zlotnick94}. Using pentameric building blocks, he enumerated all unique compositions of one or more pentamers (each arranged exactly as it would be if the entire capsid was formed). This technique, and several following publications\cite{zlotnick2005theoretical,zlotnick06}, revealed various aspects of assembly for different viruses, including rates of assembly, effect of nucleation, detection of possible kinetic traps etc. It also provided a simple tool to predict the effect of changing environment parameters and or presence of other molecules, which can be applied to measure yields under different conditions, designing conditions amenable to specific assemblies, etc.\cite{burns2009altering,Zlotnick_Mukhopadhyay_2011}. 

 % without assuming that a resolved model (from cryo electron microscopy, or x-ray crystallography) 

This paper presents an approach to score and rank conformational ensembles of capsid protein capsomers and capsid subassemblies based on a new configurational sampling and energy analysis approach. 
The sampled configurations of capsid subassemblies represents the various potential intermediate states of a fully assembled viral capsid. In other words, we recognize that the tertiary structure (fold) of individual subunits as well as binding contacts between subunits may evolve over the span of the entire assembly process, and moreover, may exist in slightly different configurations for the same subassemblies. The presence of such uncertainties imply that any binding free energy computed solely based on the structure and interfaces that exist in the final matured state of the capsid is not always accurate. Similar uncertainty quantification and uncertainty propagation methods have recently been used for single molecule models \cite{lei2014quantifying,rasheed2015quantifying} but not for combinatorial arrangments of viral capsid proteins in various capsomeric states.

In our approach, given prior knowledge (in the form of statistical distributions) of the nature of uncertainty, we can provide additional theoretical upper bounds on the distributional moments\cite{Hoeffding_1963,Azuma_1967,McDiarmid_1989} for different properties of viral capsomers, and other quantities of interest or QOI (e.g. the binding free energy). See for e.g. \cite{ rasheed2015quantifying} for such Azuma-Hoeffding bounds applied to molecular modeling with atomistic positional uncertainty captured by B-factors. Additionally, if the space of configurations is sampled such that low-discrepancy (and also low dispersion) is achieved, then a probability distribution of the QOI can be approximated with bounded error \cite{Niederreiter_1990,James_Hoogland_Kleiss_1998}. Such estimation of binding free energies, i.e. as distributions instead of single values, makes it possible to sample energy landscapes through various configurational ensembles, and analyze binding pathways in an efficient and robust way \cite{rasheed2015quantifying}. We apply our efficient low-discrepancy product space sampling technique reported in \cite{BBCZ_2015} to generate such low-discrepancy sampled ensembles of viral capsomers. The configuration space for any subassembly of the capsid is a product space of the backbone torsion angles between relatively rigid domains, as well as 3D affine transformations between each pair of neighbors. 

We also use the Wilcoxon sign rank test \cite{wilcoxon1945} to compare the computed sampled energy distribution of various capsomers. Using this we rank all possible transitions from any given subassembly, as a step towards  stochastically predicting and analyzing stable assembly pathways, and with quantified uncertainty. Finally, we use the distributions of free energy calculations to provide an analysis of the relative concentrations of intermediate subassemblies as the capsid is being formed.

%textbf{need to add a summary of the findings here??}
%Analyzing assembly pathways require the ability to quantify and rank different energetically favorable paths in terms of their likelihood, which is most often related to the binding free energy. After ensuring the uncertainty at each stage of the assembly has been quantified efficiently, we can 

% these are not quite relevant here. I could not think of any way to tie them in. Leaving them out for now. \cite{Cheng_Brooks_2013,Bahadur_Rodier_Janin_2007,bahadur2010structural,headd2007protein}

\section{Materials and Methods}
One of the major goals of this work is to develop a method for viral self-assembly pathway analysis with statistical guarantees. We consider assembly from an equilibrium perspective, where, given prior knowledge of the final assembled structure, we can uniquely determine the possible subassemblies of different sizes and the possible ways they can be associated/disassociated. This assumption, that the binding free energy of the association governs the success and yield of the reaction, is similar to the work of Zlotnick \etal\cite{zlotnick94,zlotnick2005theoretical,zlotnick06,burns2009altering,Zlotnick_Mukhopadhyay_2011}, but applies a more robust estimation of the binding energy under an uncertainty quantification framework. Additionally, we consider all possible assembly pathways starting from monomers, instead of assuming that trimers, pentamers etc.\ are the basic building blocks.

The overall methodology for this research is as follows. First, we identify all unique interfaces and unique subassemblies of specific sizes (where ``size'' is defined as the number of constituent monomers) of a given virus. Second, we sample the space of configurations for each of these subassemblies with restricted range of motion to generate an ensemble of structures in an attempt to capture the uncertainty (flexibility, random perturbations etc.) of the structure of the subassembly. Then, we compute the free energy of each sample of each subassembly to generate a distribution of the energy. Finally, we use these distributions of energies (instead of the traditional single value) to compare the stabilities of the subassemblies, compute distributions of binding free energies between subassemblies, predict likely transition pathways from one subassembly to another, etc. In the following subsections, we discuss each of these in detail.

%For this work, we have chosen to model the self-assembly process as an additive one, namely that individual {\it subunits} (single protein chains), e.g. $P_A$ and $P_B$, join together to form a complex, called a {\it subassembly}, $P_{A+B}$, which changes the energy of the system by some amount, $\Delta E(P_{A+B}) = E(P_{A+B}) - E(P_A) - E(P_B)$. Individual subassemblies will continue to form a complex with other subassemblies (or individual subunits) until the entire protein capsid is formed. The self-assembly process will follow the pathway where the $\Delta E$ is most favorable (i.e. lowest).

\subsection{Unique Subassemblies and Transitions}
Analysis of self-assembly focusing on only a predetermined set of pathways, fails to take into account subassemblies caught into energy traps, which nonetheless, may be part of a pathway which is globally favorable to the capsid as a whole \cite{hagan06,Elrad_Hagan_2008}. For this reason, we have sought to implement an exhaustive approach. We consider all possible unique subassemblies and all possible ways they can come together. 

To begin, we consider each chain to be unique. Even though the chains have the same primary structure, in most cases they exhibit minor differences in their tertiary configuration and hence it is preferable to consider each of them as unique, especially when computing binding free energies. For subassemblies involving two or more monomers, we consider them to be equivalent if and only if all three following conditions (evaluated in this order) are met: 1) they have the same number of monomers of each type, 2) they have the same number of symmetric interfaces of each type, and 3) when the entire subassemblies are aligned, the RMSD is less than a threshold.

\begin{figure}[t]
\includegraphics[width=3.5in]{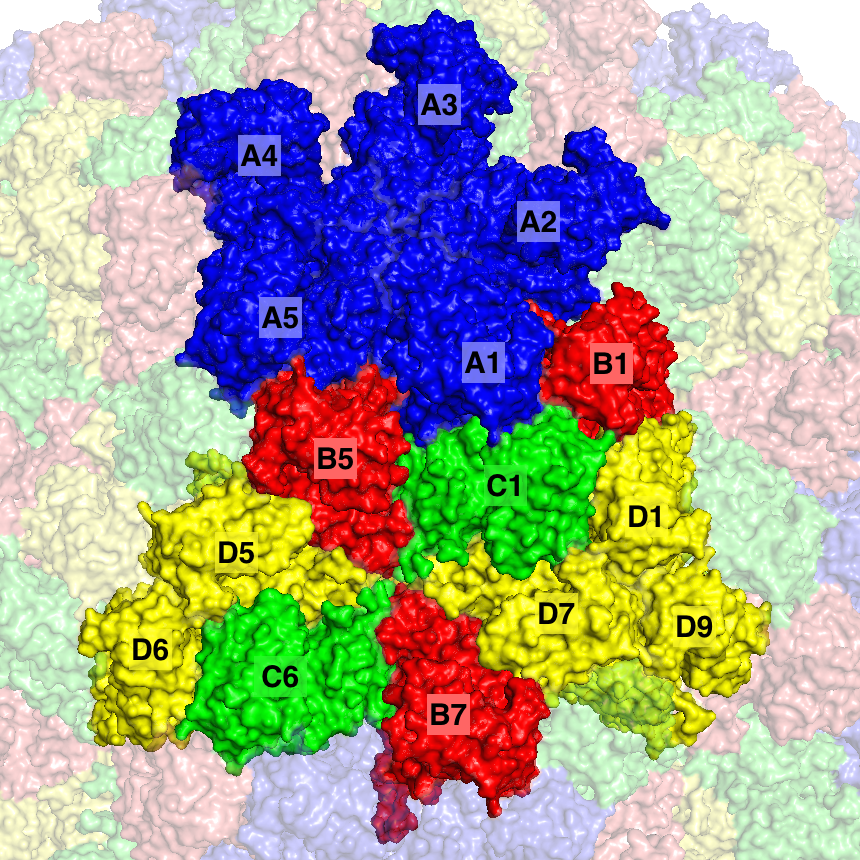}
\caption{A portion of the capsid of Nudaurelia Capensis virus (PDBID:1OHF) capsid. Labels on the capsid shows individual monomeric capsid proteins of different types A/B/C/D and at different locations and forming different local subassemblies. For instance the capsomer A1-A2-A3-A4-A5 form a pentameric configuration and contains four 5-fold interfaces of the same type. Similarly A1-B1-C1 and D1-D7-D9 are two trimers, but involve slightly different interfaces. }\label{fig:1ohf_labeled}
\end{figure}

For example, the Nudaurelia Capensis virus (PDBID:1OHF) has 240 proteins on its capsid, and has four unique monomers: A, B, C, and D. It has several unique symmetric interfaces, each appearing multiple times on the capsid. In Figure \ref{fig:1ohf_labeled}, we show a portion of capsid where each of these unique interface types are present at least once. For example, 5-fold between A1-A2, A2-A3, etc.; 6-fold between C1-B5, B5-D5, C1-D7, etc; 3-fold between A1-B1, A1-C1, B1-C1, D1-D7 etc.; and 2-fold between C1-D1, A1-B5 etc. According to our criterion, A1-B1 and A5-B5 are equivalent to each other but not equivalent to A1-B5 (violates criterion 2). A4-A5-A1-B5 and A5-A1-A2-B1 are equivalent, but A1-A2-A3-B1 are not (violates criterion 3).

We select a set of subassemblies such that no member of the set is equivalent to any other member. We ended up with 985 subassemblies involving up to 6 monomers. This set includes some of the more distinct capsomers of this capsid: the trimers (A1-B1-C1) and (D1-D7-D9), the pentamer (A1-A2-A3-A4-A5), and the hexamer (B5-C1-D7-B7-C6-D5). For this work, the number of subunits in any given subassembly was limited to 6. % for computational requirements. (including all complexes quickly becomes intractable).

We consider a transition from subassembly $\textbf{P}$ to subassembly $\textbf{Q}$ feasible if it is possible to add one monomer to $\textbf{P}$ to make it equivalent to $\textbf{Q}$. For example, for the case shown in Figure \ref{fig:1ohf_labeled}, (A1-B1-C1) is reachable from (A1-B1), (B1-C1), and (A1-C1).

%To accomplish this, we sought to solve two sub-problems:
%\begin{enumerate}
 %\item Given a set of protein subunits, $P_i$ and $P_j$ (i.e. the pentamer consisting of A1-A2-A3-A4-A5, see Figure~\ref{fig:1ohf_labeled}), determine the PDF of their binding free energy, $\Delta E(P_i+P_j)$, and
 %\item Given a set of energy probability distributions (PDFs) $\Delta E(P_i+P_j)$ and $\Delta E(P_i+P_k)$ of two subassemblies, develop a comparison ranking test to determine which subassembly is more favorable. 
%\end{enumerate}
%We will discuss each of these sections in more detail here.

\subsection{Low-Discrepancy Sampling of Subassemblies}
As mentioned before, instead of considering a static model for a subassembly, we are interested in modeling its structure as a distribution of possible structures which have minor differences, but represent the same state. One way to think of this is to consider an energy well that contains the specific subassembly and many others which are just slightly different---in such case, one should not focus on only one of them to characterize the well, but should consider the entire distribution. In this regard, we are interested in both small changes inside subunit conformation (the natural shift in structure of the protein backbone) and slight perturbations of the interface. Now we describe a parameterization of these spaces.

\begin{figure}[t]
\includegraphics[width=4in]{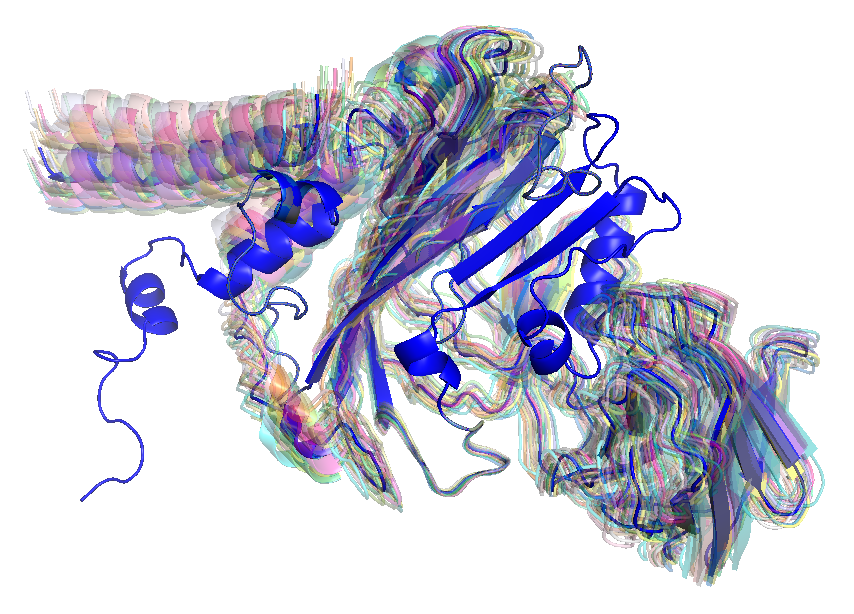}
\caption{A few sample configurations for monomer chain A, generated by sampling backbone torsion angles, shown using transparent rendering. The original configuration is rendered opaque.}
\label{fig:A_hinges}
\end{figure}

\subsubsection{Configuration space of a subassembly}
\paragraph{Internal (flexible) DOFs}
While in principle backbone torsional angles are all relevant when considering internal flexibility of a protein, for the sake of computational tractability (especially for the multitude of subassemblies we have), we applied a coarse-grained approach based on domain decomposition. Essentially, we limited the sampling space to flexible backbone torsion angles between relatively rigid subdomains. To determine the set of flexible backbone torsion angles, we used HingeProt \cite{emekli2008hingeprot} to identify {\it hinge} residues for each subunit, $P_X$, designating the corresponding $\phi$ and $\psi$ internal torsion angles of each residue as flexible (i.e. if there were $r$ hinge residues, there were a total of $2r$ rotatable bonds). This results in a configurational space equivalent to $\mathbb{R}^{2r}$. Figure~\ref{fig:A_hinges} shows an example of sampling such a space.

\begin{figure}[ht]
\includegraphics[width=4in]{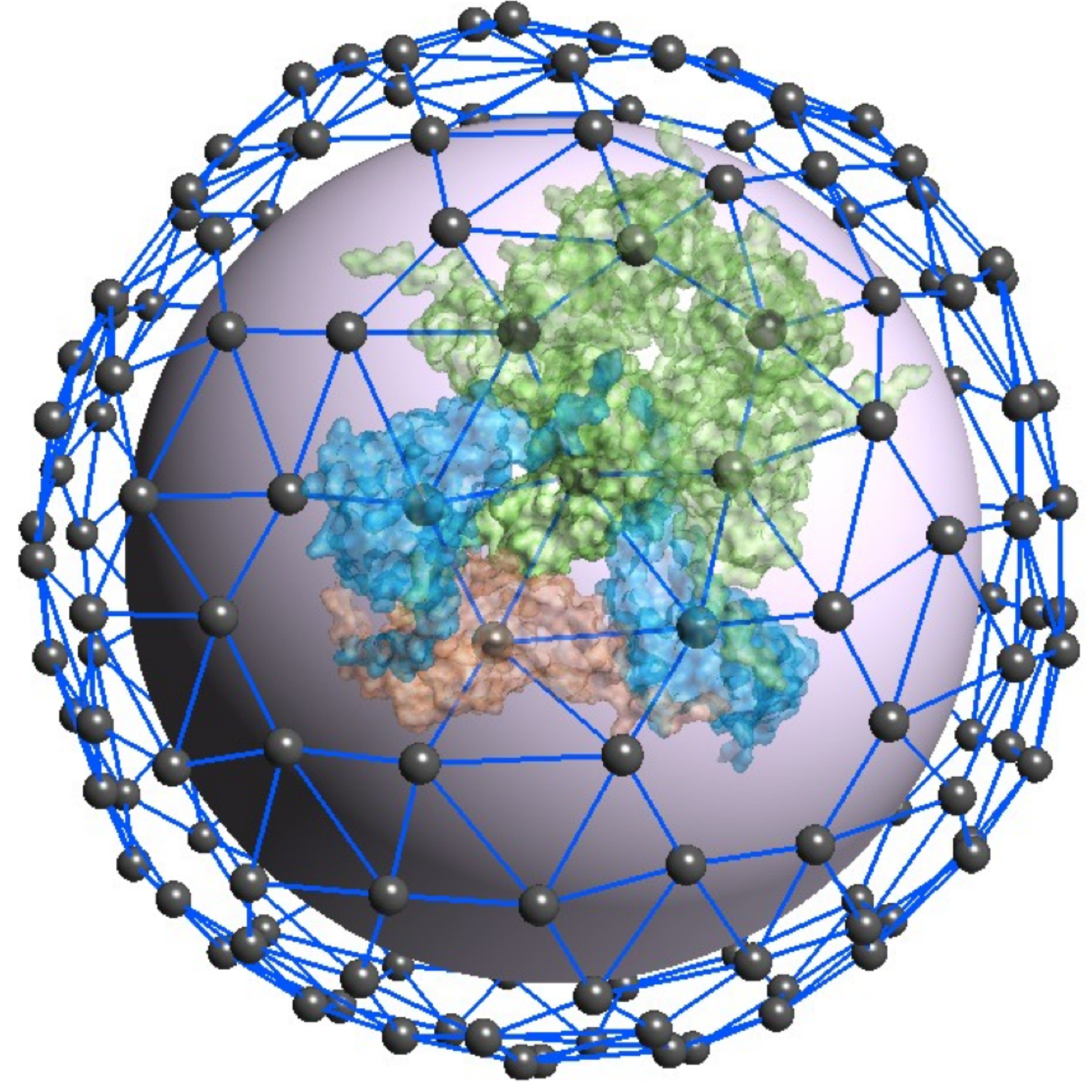}
\caption{Interface graph for 1CWP (cowpea chlorotic mottle virus). Each node on the graph represents a monomer (some of the monomers are shown). Two nodes are connected if their corresponding monomers share an interface. Any subassembly corresponds to a connected subcomponent of the graph, and the space of local perturbations of the subassembly has $6t$ DOFs, where $t$ is the number of edges in that subcomponent of the graph.}
\label{fig:dof-graph}
\end{figure}

\paragraph{External DOFs}
We parametrize the space of local affine perturbations of each pairwise interface between every pair of monomers in a subassembly using 6-DOFs, defined by three Euler angle twists and three translational shifts. Hence, for a subassembly with $t$ pairwise interfaces (edges in a graph like the one in Figure \ref{fig:dof-graph}), the space is equivalent to $\mathbb{R}^{6t}$

\subsubsection{Sampling}
Recall that we want to estimate the distribution of free energy over configurations in a local neighborhood of a given subassembly. Computing such distributions analytically over such a space is beyond the scope of our current work. What we intend to do instead is to provide an approximation of the distribution through discrete sampling of the configurational space. First, we show that if the set of samples fulfill certain conditions, then the estimated distribution approximates the correct distribution.

\paragraph{Bounded error of estimation through low-discrepancy sampling}
For a continuous function $f$ on a $d$-dimensional product space $\mathcal{I}^d$, the modulus of continuity is defined as 
$\omega(f,t) = \sup_{u,v \in \mathcal{I}^d \& {\delta}(u,v) \leq t} |f(u) - f(v)|$, where ${\delta}(u,v)$ is the distance between two configurations/samples. In other words, the value of $f$ does not change without bounds if the parameters are close. Also given a set of $N$ samples $P = \{x_1, x_2, \ldots, x_N\}$, we can define their discrepancy with respect a collection of subsets, $\mathcal{X}$, as:

\begin{equation}
 D(P, \mathcal{X}) = \max_{X\in\mathcal{X}}\left(\frac{|P\cap X|}{|P|}-\frac{\mu(X)}{\mu(\mathcal{U})}\right),
\end{equation}
where $\mu$ is the Lebesgue measure (high-dimensional volume), and $\mathcal{U}$ is the universe. Discrepancy can the be thought of as the ``evenness'' of the sample distribution. 

Now we present the following theorem:\\

\noindent
\textbf{Theorem 1: Bounded error of integral over $\mathcal{I}^d$ (adapted from Theorem 2.13 of \cite{book:neid})}~\\
If $f$ is continuous in $\mathcal{I}^d$, then, for any set of samples $P = \{x_1, x_2, \ldots, x_N\}$ such that $x_i \in \mathcal{I}^d$, we have:
\begin{equation}
\left|\int_{\mathcal{I}^d} f(u)du - \frac{1}{N} \sum^{N}_{n=1} f(x_n)\right| \leq 4\omega\left(f; \left(D_N^*(P)\right)^{1/d}\right)~\\
\end{equation}

Essentially, if one ensures that $D_N^*(P)$ is low, then the error of approximation for the integral is bounded. In our case, we want to approximate a distribution. Notice that the above theorem guarantees that if low-discrepancy sampling is performed, the cumulative distribution function (CDF), as well as the moments will be approximated with bounded error.

However, generating such low-discrepancy sampling in a high-dimensional space is nontrivial.

\paragraph{Efficient low-discrepancy sampling in high dimensional spaces}
Naive approaches in generating low-discrepancy sampling in product spaces involves sampling each degree of freedom uniformly and then combining them in all possible ways to generate samples in high dimensions. This clearly results in exponential number of samples (in terms of $d$, the dimension of the space), but also does not always guarantee low-discrepancy. Furthermore, it would be computationally intractable to achieve an acceptable level of discrepancy when using such exponential sampling over a space equivalent to $\mathbb{R}^{2r + 6t}$ for any practical values of $r$ and $t$. However, there exist efficient sampling strategies for such spaces which {\it guarantee} bounded discrepancy using only a polynomial (in the number of degrees of freedom) number of samples. 

In this article we leverage the product space sampling algorithm described by \citeauthor{BBCZ_2015}\cite{BBCZ_2015} which guarantees low-discrepancy defined over subsets of combinatorial rectangles ($\mathcal{X}$, in the definition in the previous subsection). The basic concept of the algorithm is to use a specific class of pseudo-random number generators (described by \citeauthor{GMRZ_13}\cite{GMRZ_13}) that provide strict randomness guarantees. These generators are then used to choose a polynomial-sized subset of samples out of the exponential-sized naive sample space, while still guaranteeing that the discrepancy of the samples is bounded by $d\delta$, $d$ being the number of dimensions and $\delta$ being the discrepancy in a single dimension. It was shown in previous work (see the figure from \cite{rasheed2015quantifying}) that for dimension greater than about 10, this method of pseudo-random sampling far outperformed traditional methods. As our dimension is $2r+6t$ (which is much greater than 10 for any practical values of $r$ and $t$), we have used this method here.

This technique was previously applied by \citeauthor{rasheed2015quantifying}\cite{rasheed2015quantifying} to bound the uncertainties of different proteins and complexes under large conformational shifts as well as local perturbations. It was found that even with high degrees of freedom, if the range of perturbations and flexible motions are constrained within a neighborhood, a relatively small number of samples are sufficient in providing low approximation error for the distribution of different quantities of interest (e.g.\ free energy). In this study, we generated 1000 samples for each subassembly.

\subsection{Computing Distribution of Free Energy and Binding Free Energy}
Given a set of samples with low discrepancy, we first compute the free energy for each of the samples. We use Gibbs model of free energy defined as $E = E_{bonded} + E_{vdw} + E_{coul} + G_{cav} + G_{vdw} + G_{pol} - TS$ where $E_{bonded}$ is bonded energy terms representing the energy required to move away from ideal bond lengths, angles, etc., $E_{vdw}$ is the internal van der Waals energy, $E_{coul}$ is electrostatic interaction energy, $G_{cav}$ is approximated using the volume of the protein and the exposed surface area, $G_{vdw}$ is the Van der Waals interaction between exposed atoms and solvent atoms, $G_{pol} $ is the polarization energy (we used Generalized Born approximation), $T$ is the temperature and $S$ is the entropy. In this article, we disregard the effect of $S$.

We used MolEnergy \cite{BCR11,BZ10MS} to compute the molecular surface, area, volume etc. and a fast GPU-accelarated algorithm, PMEOPA \cite{cha2015accelerated} for computing the van der Waals, Coulombic and polarization energies. The accuracy of these techniques were established by \citeauthor{cha2015accelerated} \cite{cha2015accelerated} by comparison with AMBER \cite{Amber}. 

While it is trivial to compute binding free energies for static cases simply as the difference of the total free energies before and after binding, it is nontrivial when the input is in the form of distributions, instead. The general idea, however, is still the same, and binding free energies can be computed as an extension of the single-quantity methodology. First, we define the binding free energies for static cases, as follows: Given a complex or assembly, $\mathbf{P}$, consisting of a set, $\mathbb{S}$, of individual chains, we express the binding free energy of $\mathbf{P}$ as:

\begin{equation}
\Delta E (\mathbf{P}) = E(\mathbf{P}) - \sum_{C \in \mathbb{S}} E(C).
\end{equation}

Now, since each of the components in the above equation is a distribution instead of a scalar, we use a probabilistic definition for the distribution of $\Delta E (\mathbf{P})$. The distribution is approximated based on a collection of 1000 observations. Each observation randomly selects a value from the distribution $E(\mathbf{P})$, and from each distribution $E(C)$ such that $C \in \mathbb{S}$.

\subsection{Comparing Distributions to Rank Assemblies and Transitions}
Analyzing assembly pathways requires the ability to quantify and rank different paths in terms of their likelihood, which is most often related to the binding free energy. Since we are dealing with distributions of such energies (rather than simple scalars), we need a slightly involved technique to compare and rank such distributions. One possible approach is to compare the moments (mean and variance, for instance), but this loses information such as whether the distribution is unimodal or bimodal. The pairwise Wilcoxon signed-rank test \cite{wilcoxon1945} uses the entire distribution and provides a way to generate a total ordering among a set of distributions. 

For any pair of distributions, $X$ and $Y$ with $N$ points, the Wilcoxon statistic, $W(X,Y)$, is computed as follows:
\begin{enumerate}
\item Let $d_i=|x_i-y_i|$, $i\in1\ldots N$, be the absolute difference between two random, independent draws, $x_i\in X$ and $y_i\in Y$; let $\sigma_i$ be the sign of $x_i-y_i$
\item Order each $d_i$ from smallest to largest, and let $R_i$ be the rank from this ordering
\item Calculate the Wilcoxon signed-rank statistic, $W(X,Y)$, as:
\begin{equation}
W(X,Y) = \sum_{i=1}^N \sigma_i \cdot R_i
\end{equation}
\end{enumerate}
It is easy to see that this statistic is symmetric, and that $W(X,Y)\approx -W(Y,X)$. 

This can be extended to multiple distributions, $X_1,\cdots,X_m$, as follows:
\begin{equation}
W(X_i | X_1,\cdots,X_m) = \sum_{j\ne i}^m W(X_i, X_j)
\end{equation}
For distributions of energy, a lower value of $W$ is more favorable, thus the most optimal distribution will have the most negative $W$ statistic. Worse (less favorable) distributions will have increasingly positive $W$ statistic.

\section{Results and Discussion}
%Predicting the self-assembly process for viral capsids can be a difficult process, especially when the number of proteins is large. One of the major problems is the underlying uncertainty at each step of the assembly. As the assembly continues, this error accumulates, and if left unchecked, can provide a solution that is uninformative. In this research, we have attempted to provide a template for a viral assembly with quantified uncertainty. In this section, we will highlight some of the major findings of our approach.

\subsection{Sufficiency of Sampling}
% Generate statistics that shows that 1000 samples is enough--reaches saturation
The first and most important question is to determine whether the number of samples we have generated (1000) is sufficient for accurate methods of moments calculations. In previous work \cite{rasheed2015quantifying}, we used an incremental sampling approach to determine when our sampling was sufficient enough to obtain confident representative distributions. For all of the single proteins in the study, less than 400 samples was sufficient to provide high confidence in error bounds. As capsid protein complexes consist of many different protein subunits, the uncertainty in a single sample can propagate and influence the stability of the complex as a whole. It is important to ensure we have achieved sufficient sampling.

\hide{
\begin{figure}
\includegraphics[width=3in]{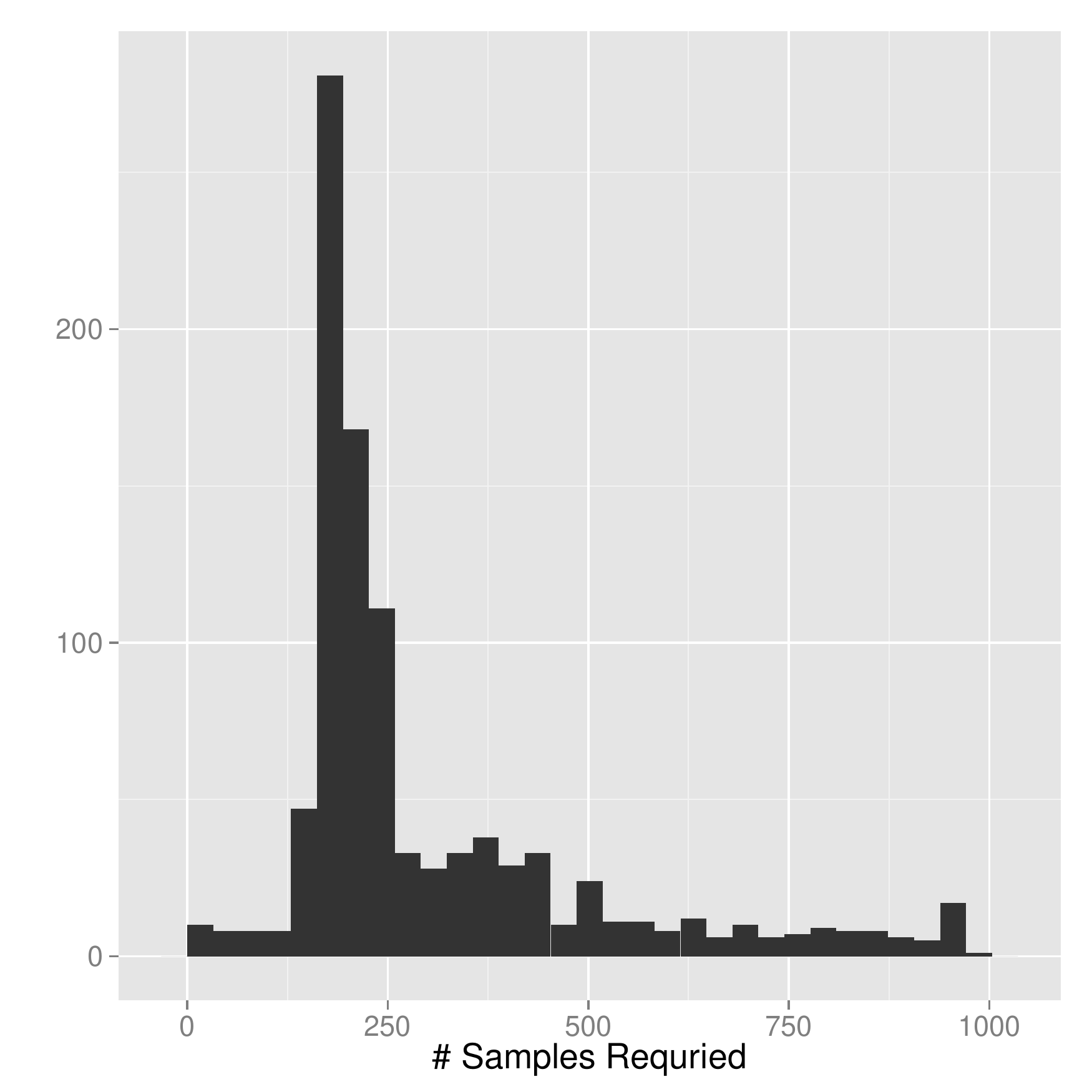}
\caption{Number of samples to reach saturation against Chernoff-like values. The subassembly that required the greatest number of samples was A1-B5-C1-D1-D5, requiring 973 samples. Less than 3\% of the subassemblies required more than 900 samples.}\label{fig:nsamps}
\end{figure}
}

% latex table generated in R 3.2.1 by xtable 1.8-0 package
% Sun Mar 13 23:52:02 2016
\begin{table}
%\centering
\begin{tabular}{cr}
\caption{Number of samples needed to reach saturation across all samples. Number of samples defined by Chernoff-like bounds for incremental sampling approach. Most of the samples (86\%) require less than 500 samples; 79\% required less that 400.}\label{tab:nsamp}

%  \hline
  \# samples needed for saturation & \% of subassemblies \\ 
  \hline
  $<$100 & 2.6 \\
  100-200 & 39.9 \\ 
  200-300 & 26.0 \\ 
  300-400 & 10.3 \\ 
  400-500 & 7.7 \\ 
  500-600 & 3.3 \\ 
  600-700 & 3.2 \\ 
  700-800 & 2.3 \\ 
  900-900 & 2.2 \\ 
  900-950 & 1.2 \\ 
  950-975 & 1.2 \\ 
  >975 & 0.0\\
   \hline
   \hline
\end{tabular}
\end{table}

%Figure~\ref{fig:nsamps} shows the distribution of 
Table~\ref{tab:nsamp} shows the required number of samples across all protein complexes. Most of the subassemblies (86\%) required less than 500 samples before saturation was reached. Only 2\% of the subassemblies required more than 900 samples, and the most unstable subassembly was A1-B5-C1-D1-D5, requiring 973 samples. (Upon closer inspection, this subassembly had several outliers that were likely affecting the results.) In our previous study, we showed that even an incremental additive procedure (samples were added until saturation had been reached) with as few as 10 additional samples provided an upper bound on the number of samples needed, so we can safely conclude that the number of samples was sufficient to obtain accurate representative distributions.

\subsection{Statistical Distribution for Individual Subassemblies}
\begin{figure}
\includegraphics[width=2in]{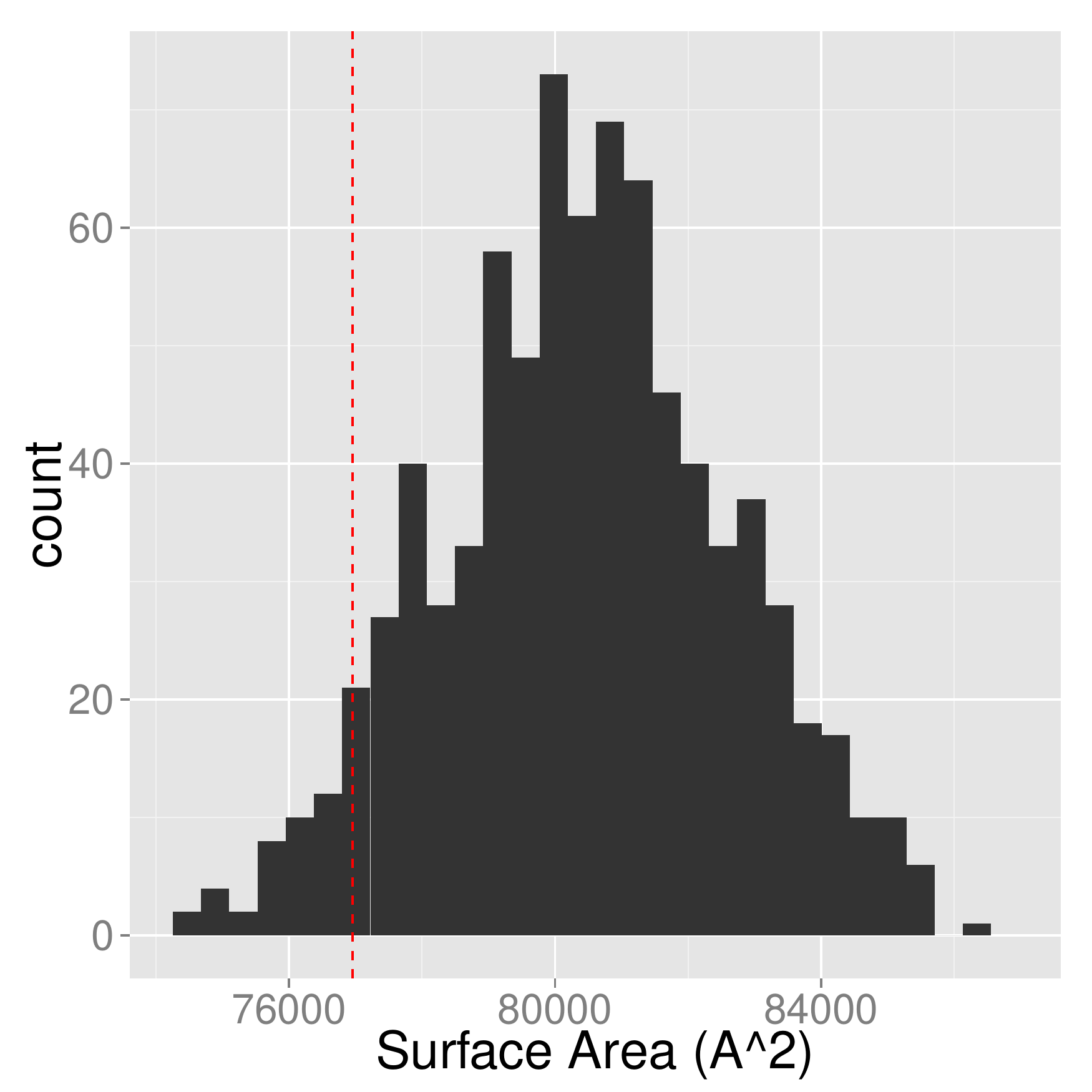}
\includegraphics[width=2in]{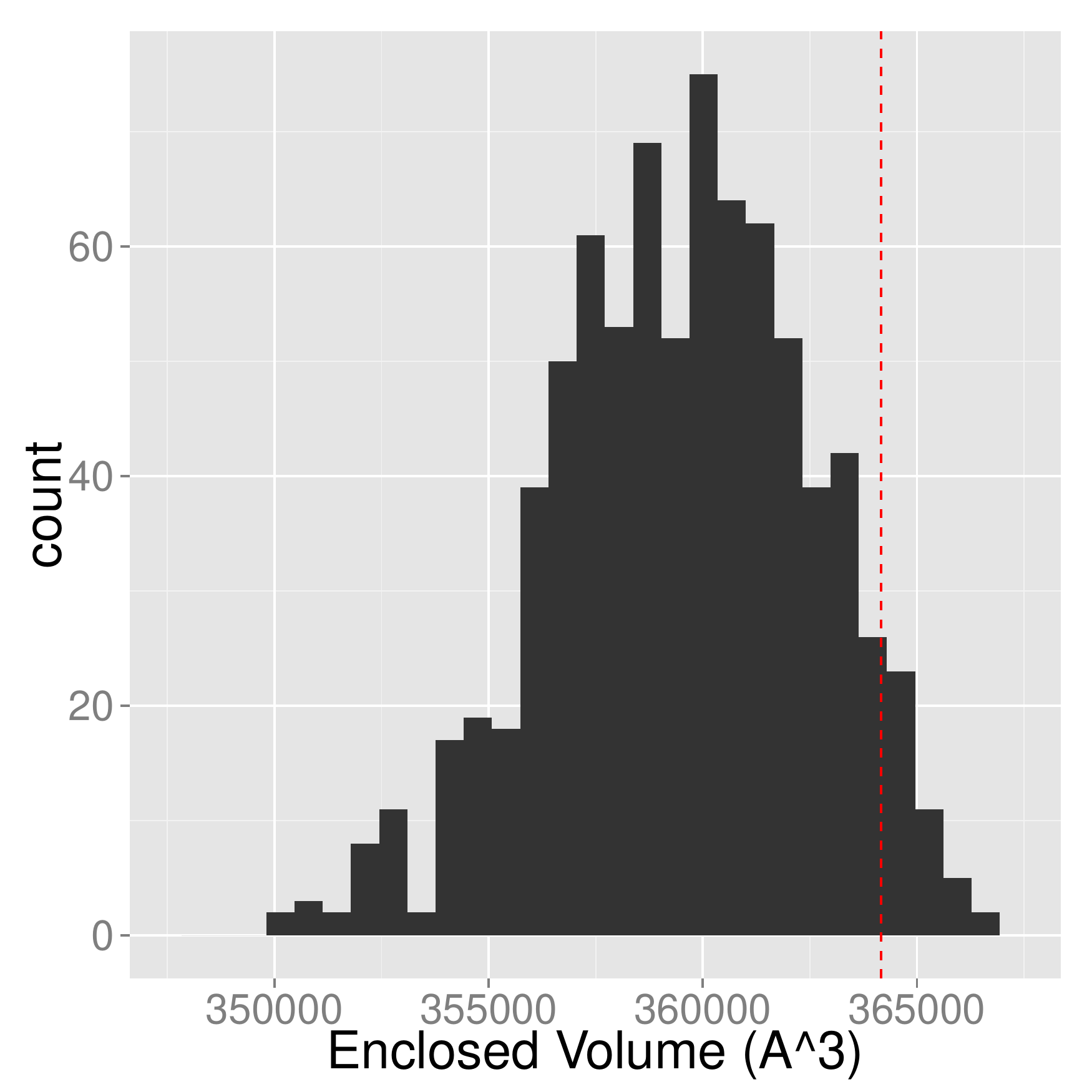}\\
\includegraphics[width=2in]{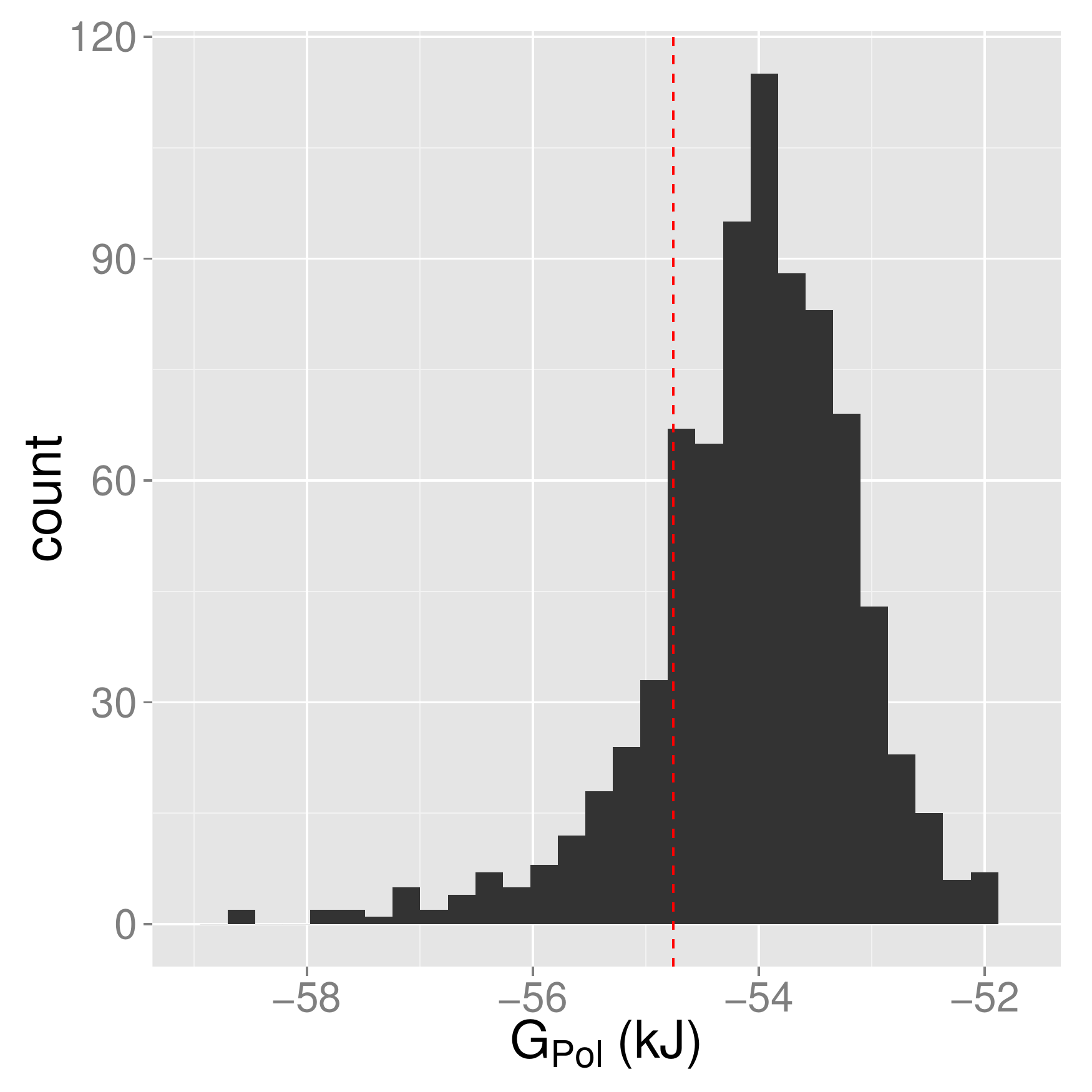}
\includegraphics[width=2in]{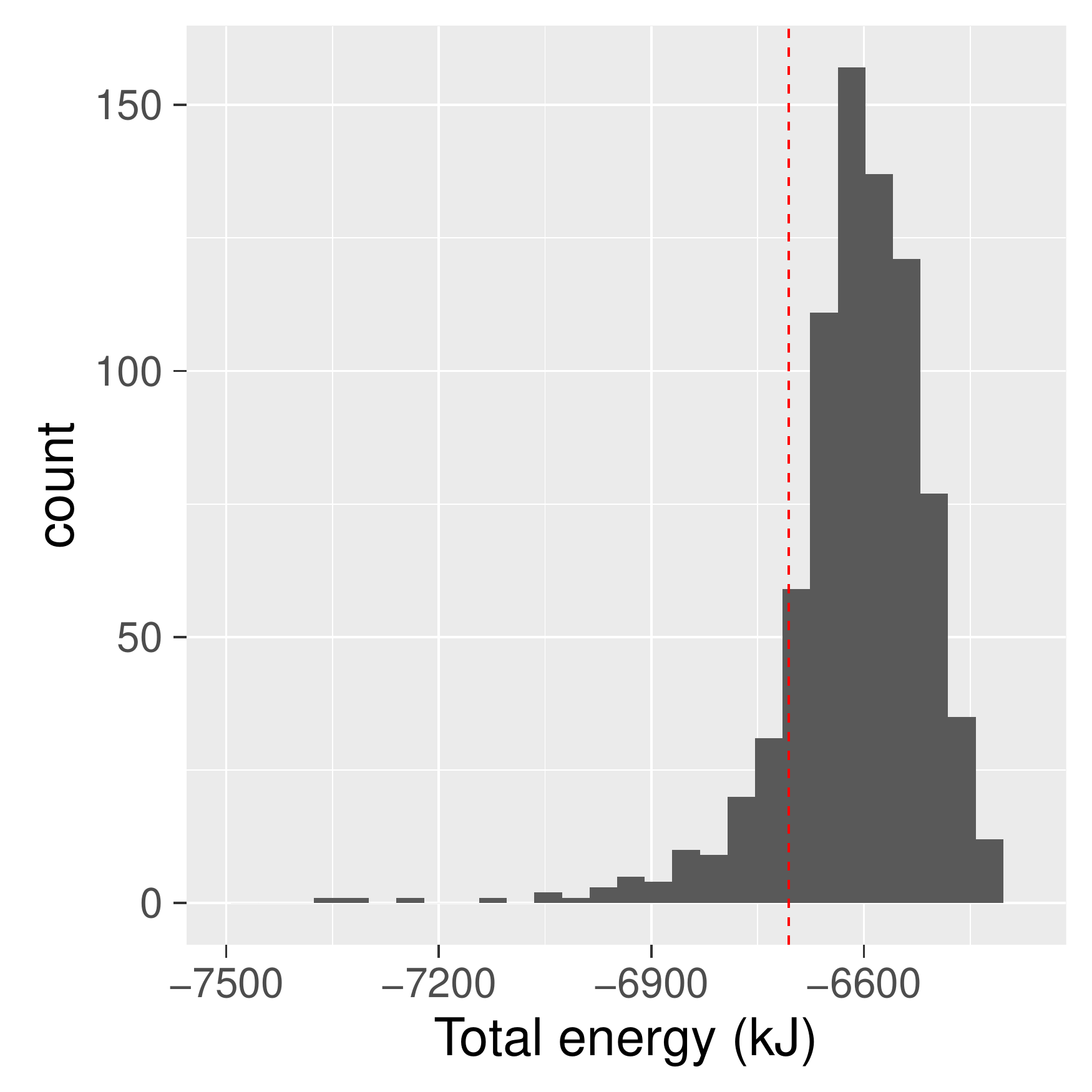}
\caption{Histogram plots of exposed surface area, enclosed volume, $G_{pol}$ and total energy (GBSA model) for all samples of the pentamer A1-A2-A3-A4-A5. Dotted vertical lines are the quantity computed on the un-sampled molecule.}\label{fig:pent_hist}
\end{figure}

% Shows plots of energy values for pentamer?
%  - Change label from GB Polar to G_{pol}
We computed the following quantities for each sample of each subassembly: exposed surface area, enclosed volume, LJ and Coulombic potentials, the solute-solvent polarization energy ($G_{pol}$), total free energy (GBSA model) and delta energy ($\Delta G$). Sufficiency of the sampling guarantees that the distributions of each of these properties are acceptably accurate. Figure~\ref{fig:pent_hist} shows the distributions of calculated surface area, exposed volume, energy and $G_{pol}$ for the pentamer, A1-A2-A3-A4-A5. As can be seen in these plots, minor perturbations in internal angles and interface contacts can have large effect on all computed quantities. Some of these changes are intuitive (small changes in internal angles have a large effect on exposed surface area, as seen by the large second moment of the PDF), but computing the quantities on all samples provides an accurate measurement as to {\it how much} they can change. Additionally, while most distributions are relatively well-behaved (approximately Gaussian with only one peak), for a small number of subassemblies (especially those with potentially few contacts), the PDF is bimodal, providing additional insight into the stability of the complex (see Supplemental~\Cref{fig:s-b1-b5-b7-c1,fig:s-b5-d7,fig:s-b1-b5}).

%
% Why are we doing these plots? Original is not always mean--discuss stdev
% Histogram of z-scores for original statistics
The second observation from these plots is one major motivation for using distributions of quantities instead of single values. In Figure~\ref{fig:pent_hist}, the dotted red line shows the quantity computed on the non-perturbed subassembly. For the pentamer, the original values vary wildly from the distribution mean. In fact, a histogram of Z-scores for all subassemblies shows that most of them differ greatly from the sampled mean, with many being more than 2 standard deviations away. In addition, subassemblies with a larger number of subunits do not necessarily have a higher variance, so correctly accounting for propagated uncertainty requires good, low-discrepancy samples.
\hide{
\begin{figure}
 \includegraphics[width=5in]{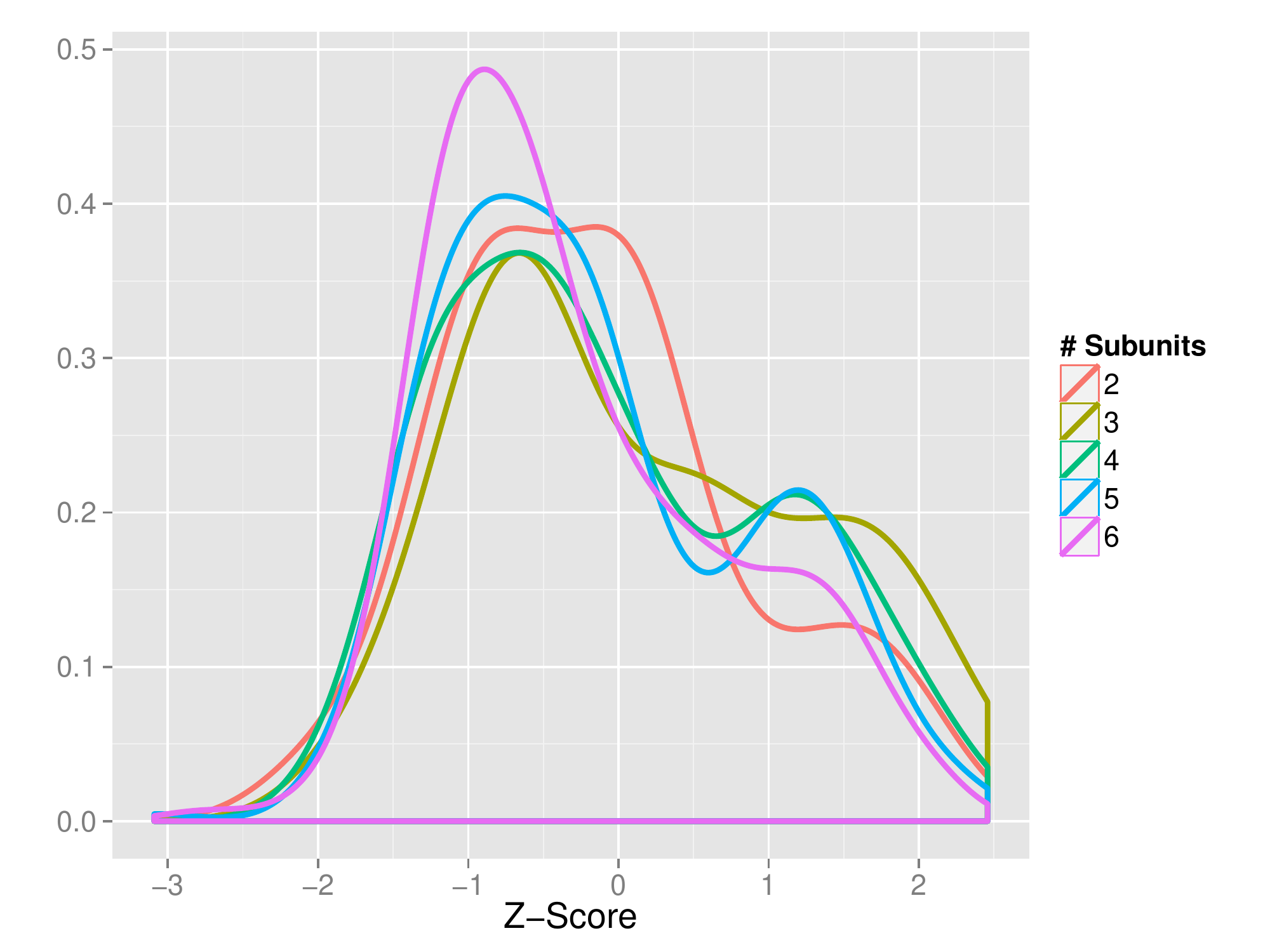}
 \caption{Distributions of Z-scores for total energy (GBSA model) computed on the native structures, across all assemblies. For many subassemblies, the Z-score is insignificant, but for many (25 subassemblies, or approximately 3\%), the Z-score is greater than 2.}
\label{fig:z-scores}
\end{figure}}

\subsection{Comparing Capsomers}
% Create uniqueness check, then only include the dimers, trimers that are unique
% just E plots:
%  - Plot of density for all 2's (just E)
%  - Plot of all 3's
Given the distributions of a number of subassemblies for a specific property, we can compare them using the Wilcoxon signed rank test and generate a total ordering. This is especially useful in gaining insights (with quantified uncertainty bounds) into the relative binding affinities or stabilities of different capsomers. 

\begin{figure}
 \includegraphics[width=0.48\linewidth]{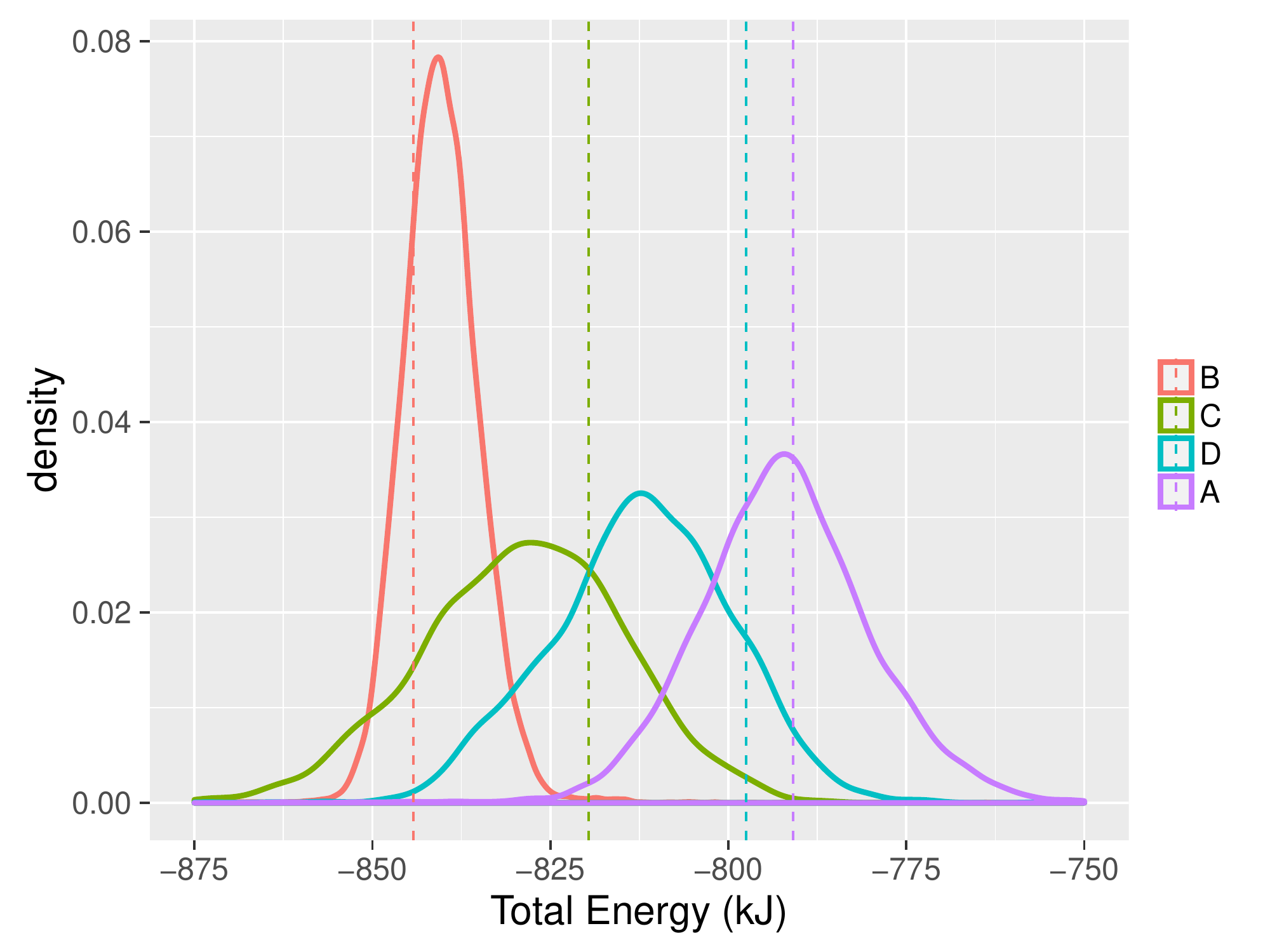}
 \includegraphics[width=0.48\linewidth]{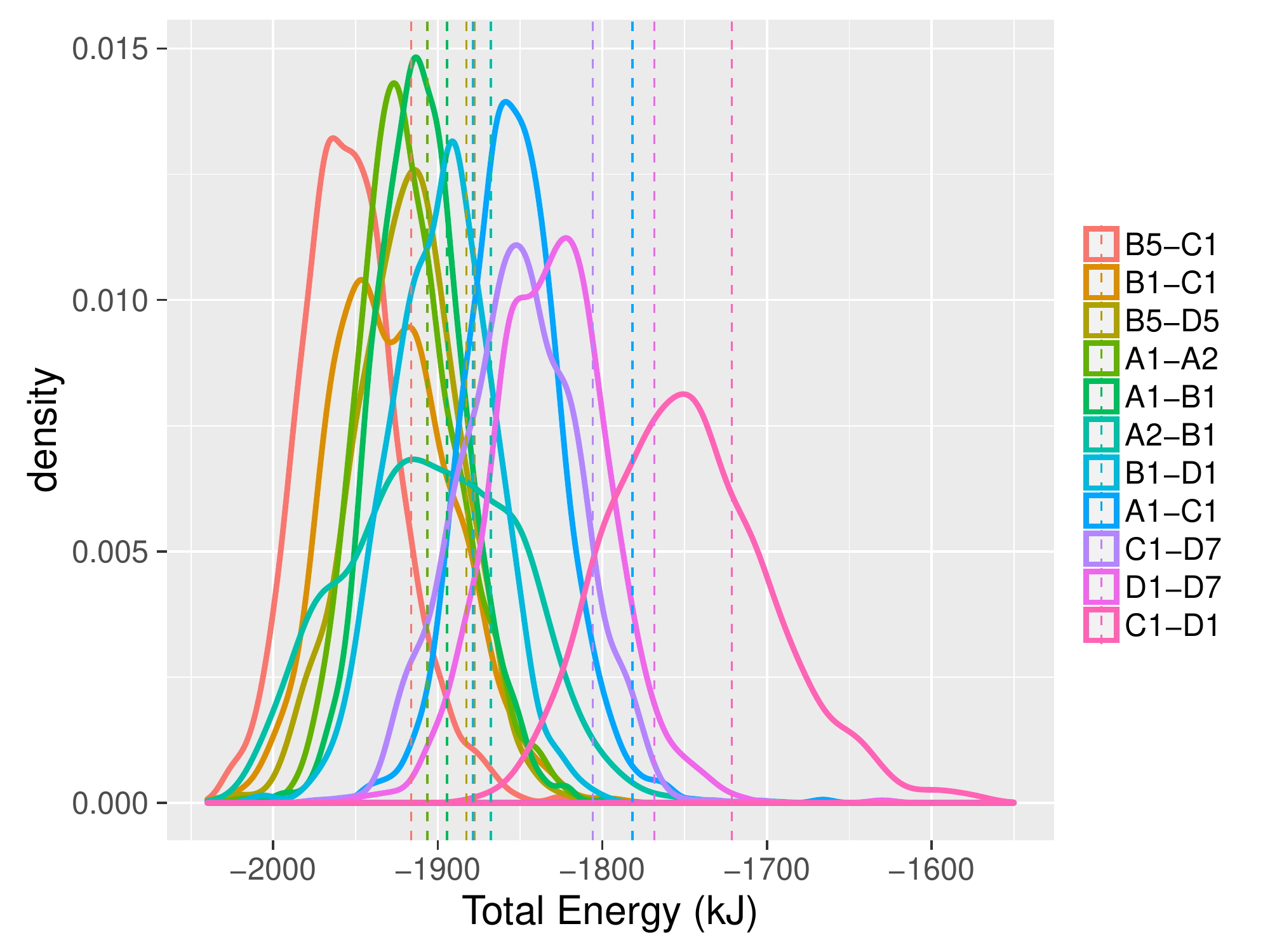}
 \caption{Density plots of distribution of all energy values for all single subunits (left) and subassemblies of size 2 (right). Legend is ranked according to the Wilcoxon signed-rank test (top is best), as reported above.}\label{fig:mon-dim-wilcox}
\end{figure}

\begin{figure}
 \includegraphics[width=5in]{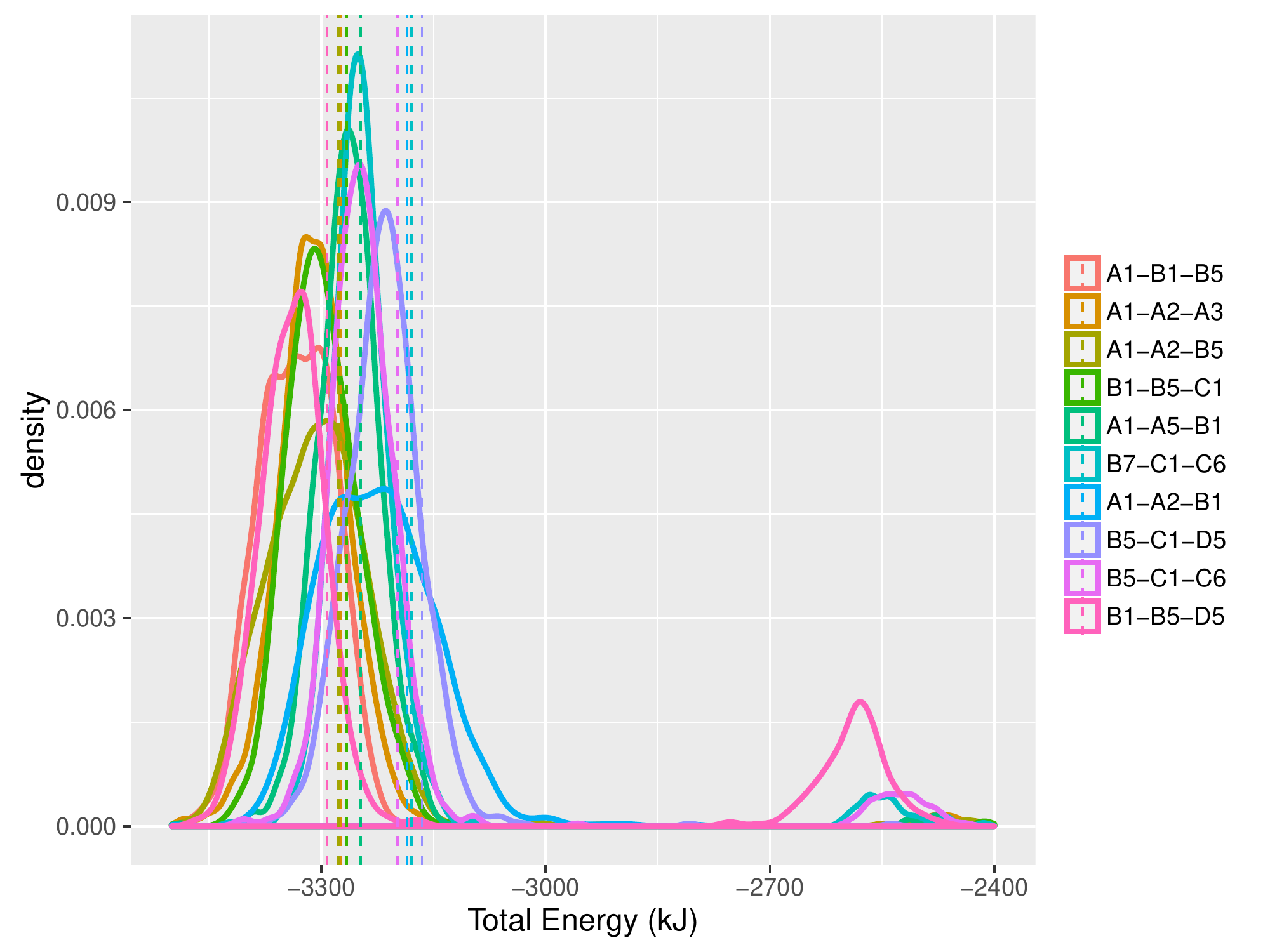}
 \caption{Density plots of distribution of energy values for {\it top 10} subassemblies of size 3. Legend is ranked according to the Wilcoxon signed-rank test (top is best), as reported above.}\label{fig:tri-wilcox}
\end{figure}

\begin{figure}
  \includegraphics[width=0.4\linewidth]{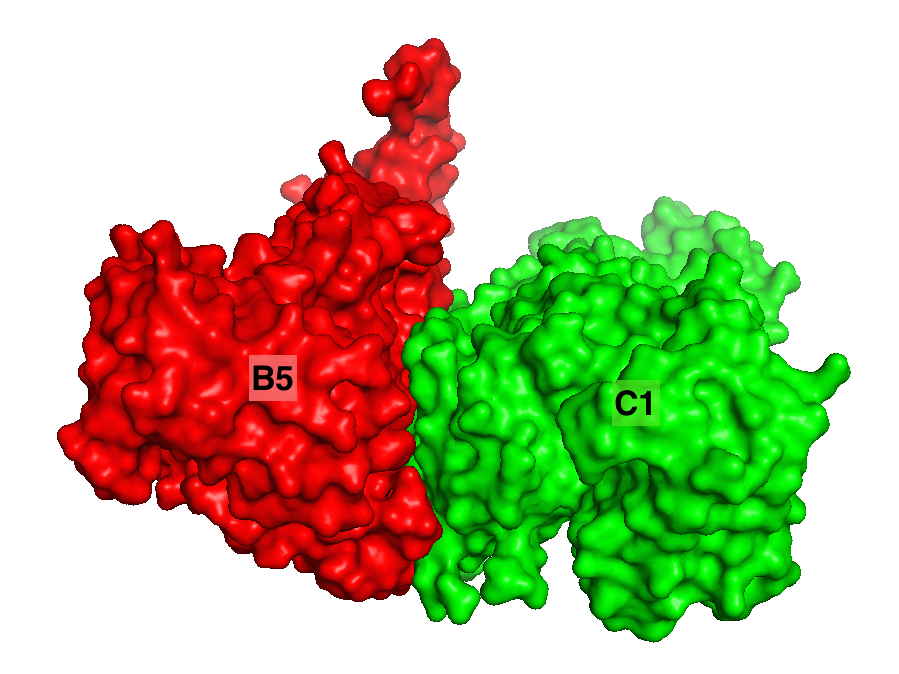}
  \includegraphics[width=0.4\linewidth]{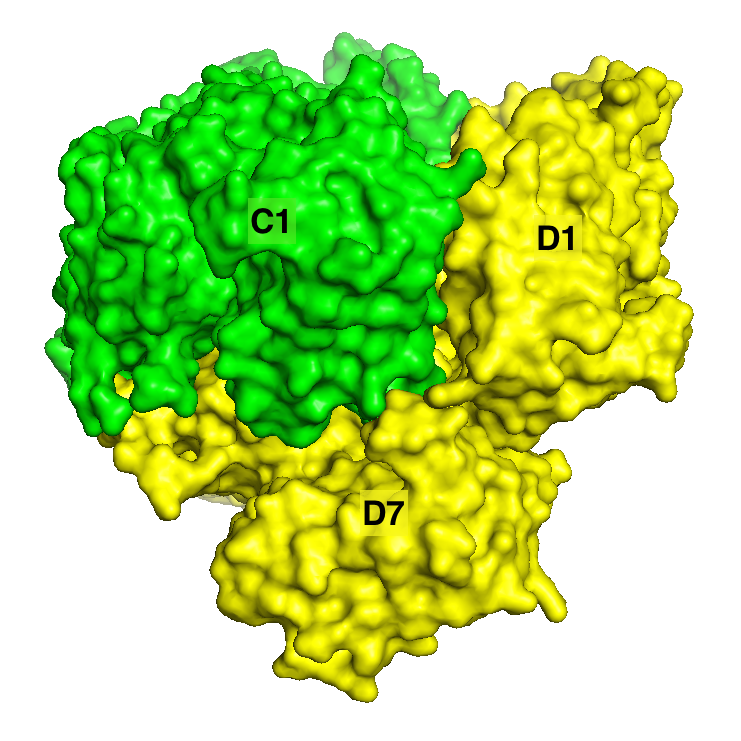}
  \caption{Labeled surface representation of B5-C1 (left), and C1-D1-D7 (right). B5-C1 and C1-D1-D7 are the most stable 2- and 3-subunit capsomers.}\label{fig:stab-surfaces}
\end{figure}

Figures~\ref{fig:mon-dim-wilcox} and \ref{fig:tri-wilcox} show the distribution of total energy for single subunits and dimer subassemblies, and the top 10 subassemblies of size 3, respectively (see Figure~\ref{fig:stab-surfaces} for a surface representation of these complexes). According to this test, the most stable subunit is B, the most stable subassembly of size 2 is the B5-C1 dimer, and the most stable subassembly of size 3 is the A1-B1-B5 complex. The least stable complexes are A, the C1-D1 dimer, and the C1-D1-D7 complex (not pictured).

\begin{figure}
\includegraphics[width=5in]{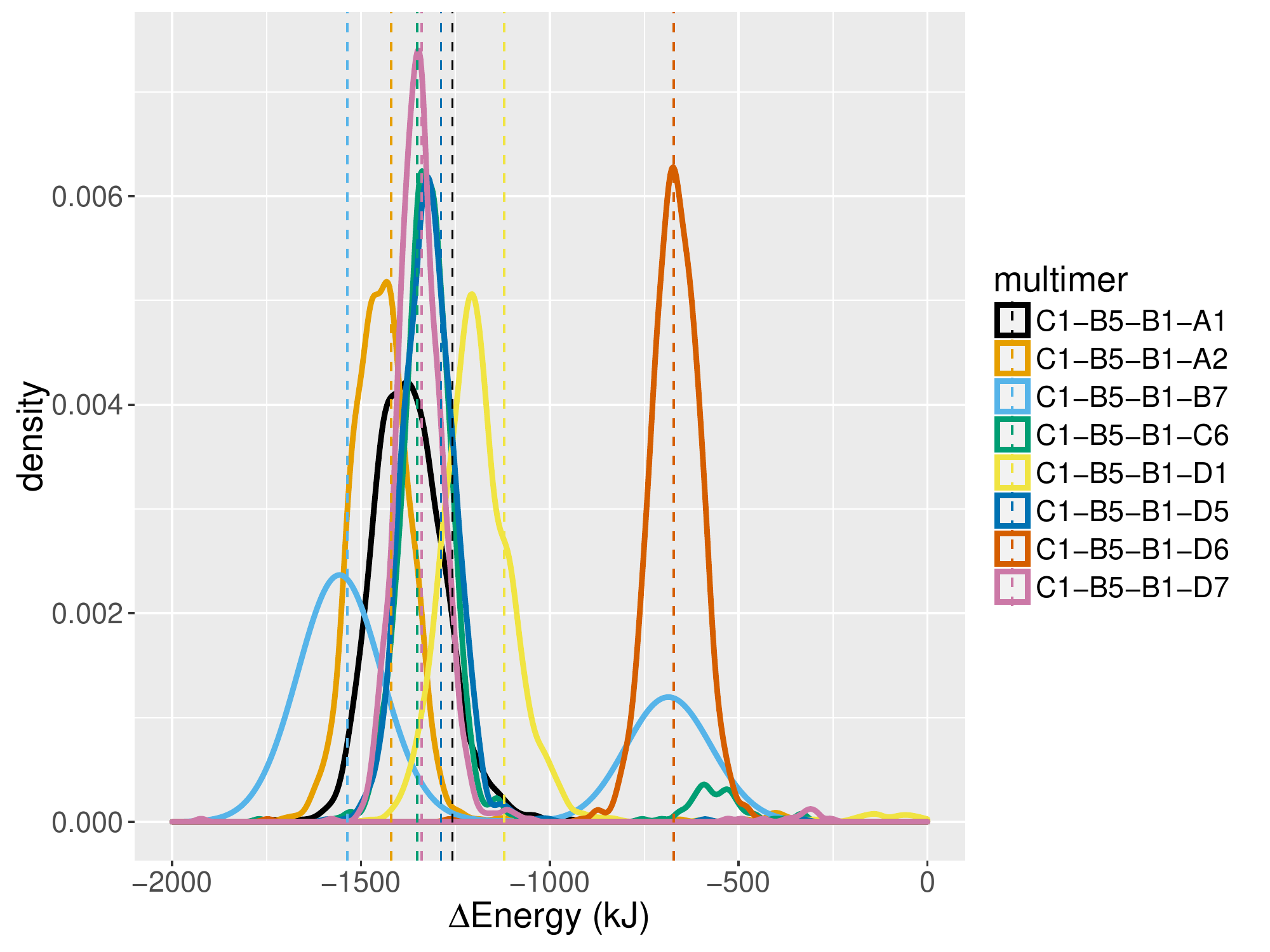}
\caption{Density plots of distribution of energy values for all possible capsomers starting with C1-B5-B1. Legend is ranked according to the Wilcoxon signed-rank test (top is best), as reported above. Dotted vertical lines show the value computed on only the native structure.}\label{fig:bad-wilcoxon}
\end{figure}

% Rank these by Wilcoxon. Say ``this dimer is the most stable'' or ``this trimer is the most stable''
We also applied the Wilcoxon signed rank test to rank all possible transitions from any given subassembly, which is a crucial step in predicting/analyzing the assembly pathway. For example, Figure~\ref{fig:bad-wilcoxon} shows the possible state transitions starting at C1-B5-B1. If only the native configurations were used, one would have reached the conclusion that adding B7 would be the best transition. However, this is a spurious result since the interface between B7 and B5 have low contact area and is only stable if D7 was also present (see Supplementary Figure~\ref{fig:s-surfs}). That sensitivity is exposed through sampling the local configuration space which resulted in some configurations with tighter binding sites and others that pulled them apart (apparent from the bimodal nature of the distribution). Our method successfully accounted for this uncertainty, and the Wilcoxon test is robust enough to determine that it is not the best possible transition.

\subsection{Transitions}
\begin{figure}
 \includegraphics[width=0.48\linewidth]{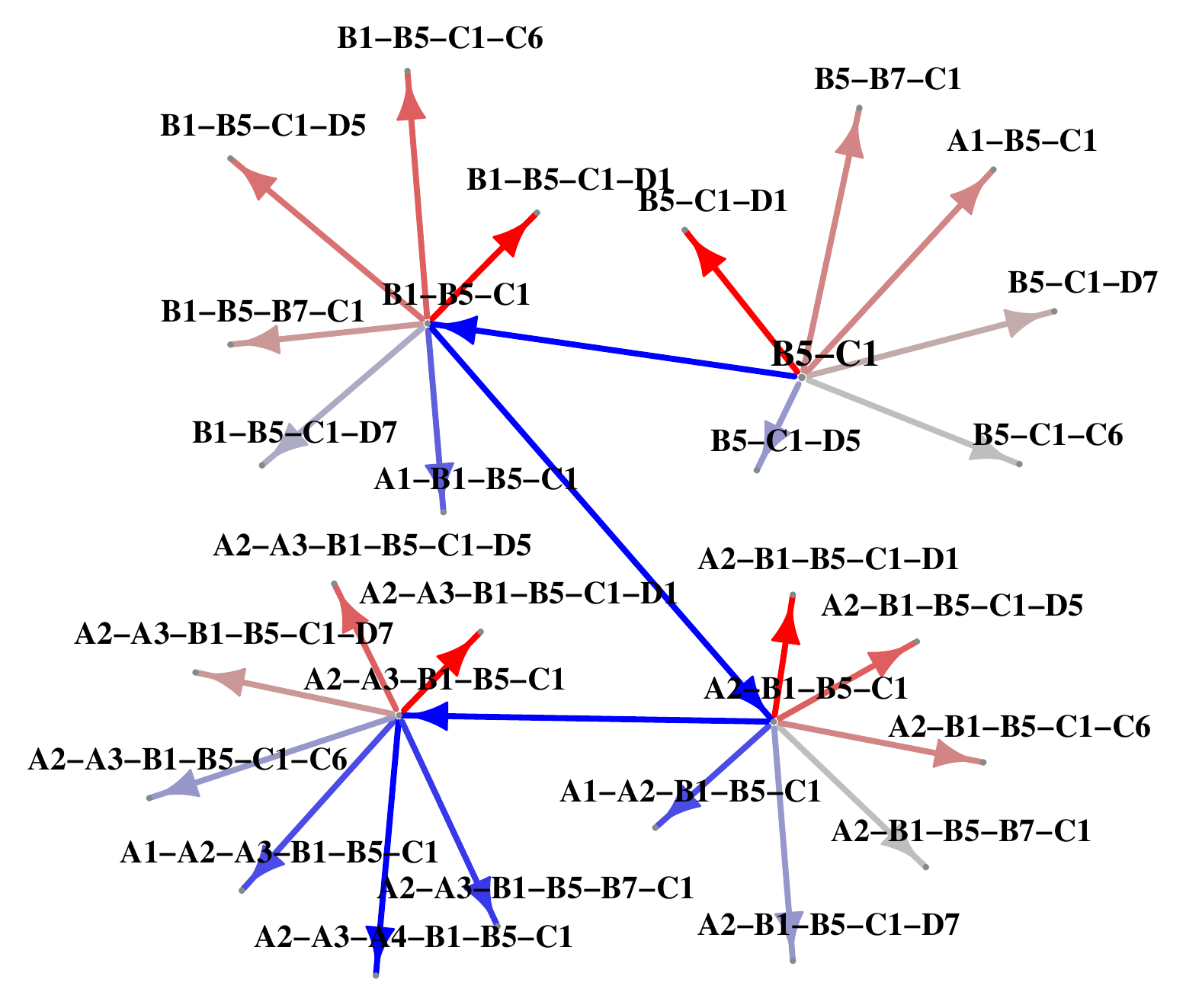}
 \includegraphics[width=0.48\linewidth]{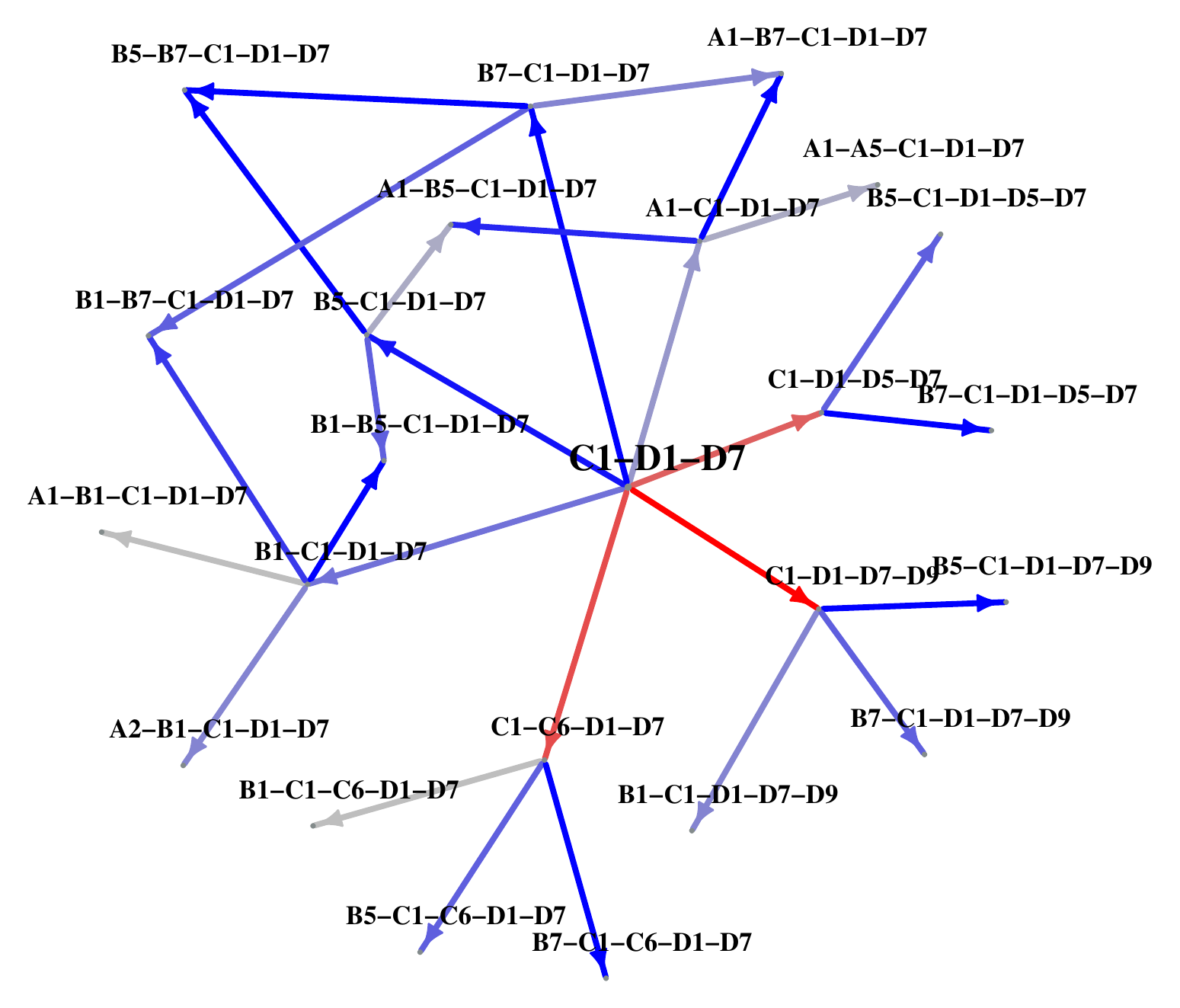}
 \caption{Network graph of possible subassemblies formed when starting from most likely starting points (left: B5-C1 dimer, and right: C1-D1-D7 complex). Color ranges from red (low Wilcoxon score) to grey to blue (high Wilcoxon score). Potential sinks are identified by nodes that have many incoming blue edges, such as A2-B1-B5 on the dimer graph.  D1-D7 complex only shows the most likely pathway; C1-D1-D7 graph has low-scoring subassemblies removed from the graph for clarity. For more network graphs of individual subunits, please see Supplemental Figures.}\label{fig:transition-graphs}
\end{figure}

% Pick the most stable dimer and most stable trimer and fan out from here
%  - Network graph of the next step
%    - one for best dimer + 1
%    - one for best trimer + 1
Finally, after ranking each transition based off its Wilcoxon score, we construct a complete transition graph showing all possible pathways leading from monomeric subassemblies to the largest subassemblies. Since the entire graph is too large to visually inspect, we present snippets of it here. Figure~\ref{fig:transition-graphs} shows transition sub-graphs starting with subunits B5-C1 and C1-D1-D7 respectively. Note that these two were determined to be very stable states according to our analysis presented in the previous subsection. Starting with (C1-B5), the most likely pathway is (C1-B5)$\rightarrow$B1$\rightarrow$A2$\rightarrow$A3$\rightarrow$A4 (Figure~\ref{fig:transition-graphs}, left). For an assembly process starting with subassembly A1-A2, the most likely pathway is (A1-A2)$\rightarrow$A3$\rightarrow$A4$\rightarrow$B5$\rightarrow$B1 (Figure~\ref{fig:transition-graphs}, right). This kind of figure can provide a visual method for observing likely state transitions. States that are very likely (such as A2-A3-B1-B5-C1, the last star node on the left graph) have many highly-weighted incoming edges (these are "sink" states). States that are not so good can only be reached via poor transitions (such as A1-A5-C1-D1-D7, in the right graph).

\begin{figure}
  \includegraphics[width=0.48\linewidth]{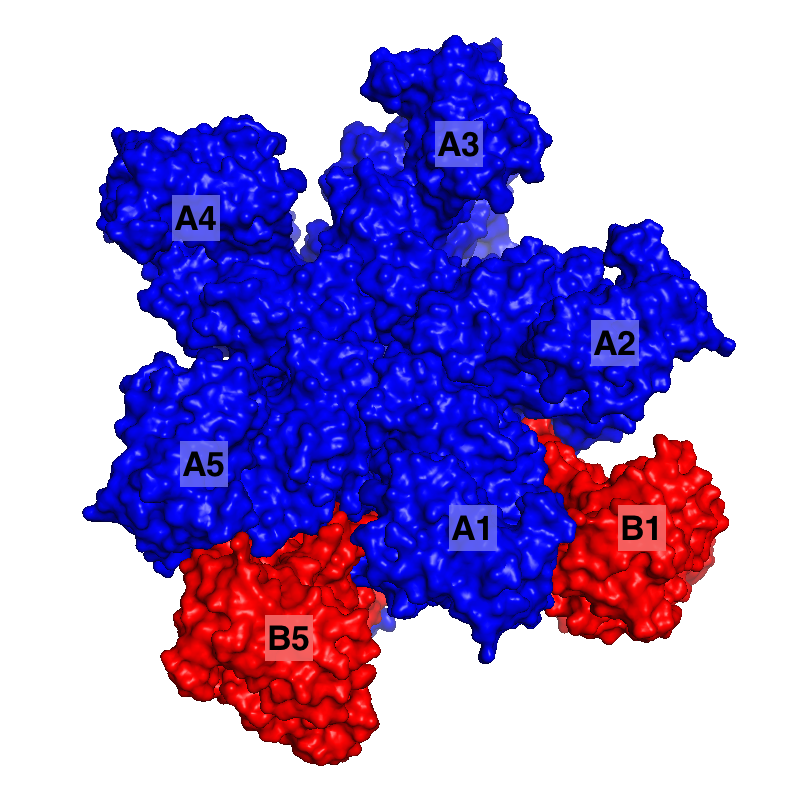}
  \includegraphics[width=0.48\linewidth]{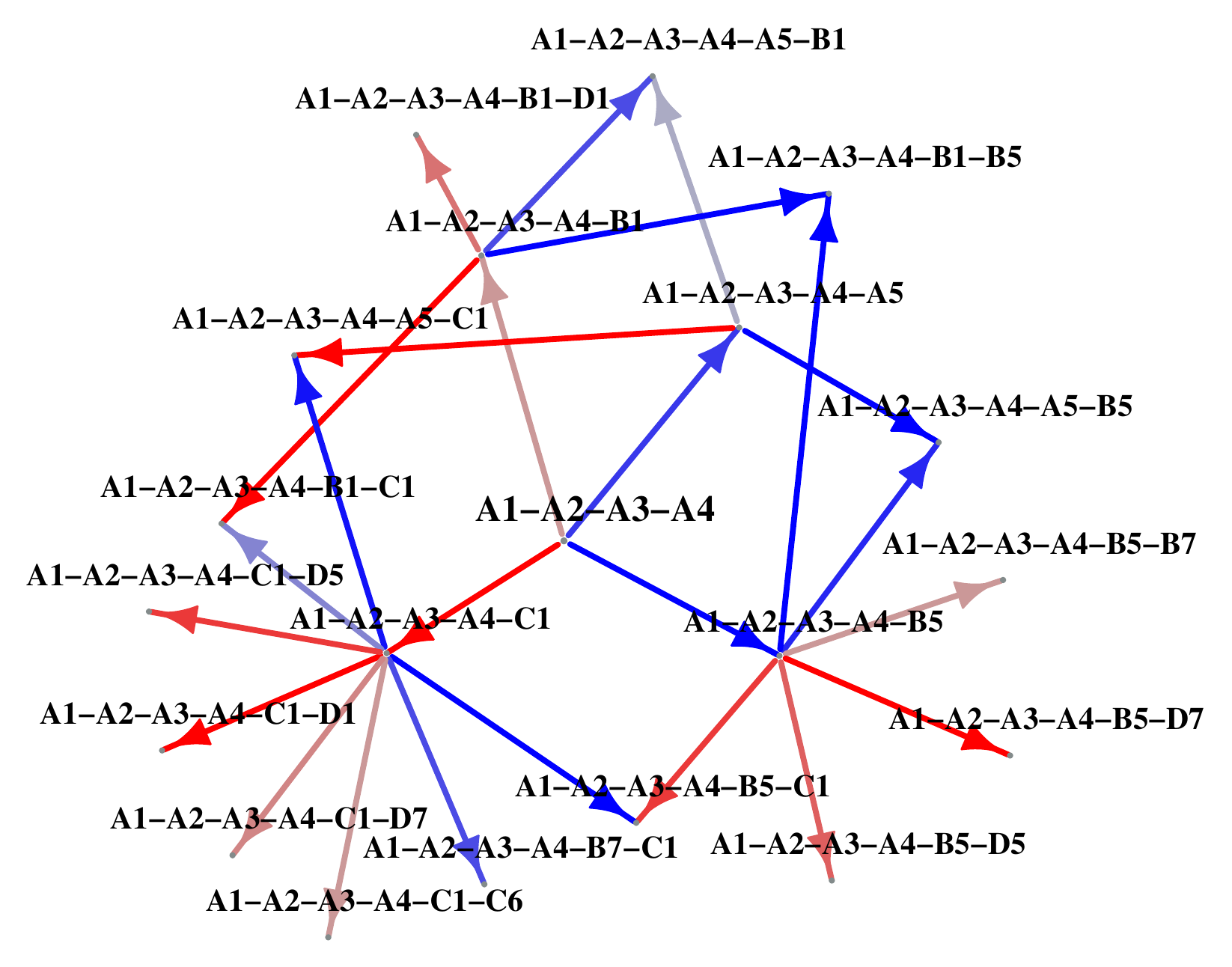}
  \caption{Labeled surface representation of A1-A2-A3-A4-A5-B1-B5 (left), and network graph of possible subassemblies formed when starting from A1-A2-A3-A4. Color ranges from red (low Wilcoxon score) to grey to blue (high Wilcoxon score). According to the Wilcoxon test and our results, the final piece of the pentamer does not form until after B1 and B5 have been added.}\label{fig:stab-pent}
\end{figure}

% Finally, show the graph (subgraph). Mention that complete graph is in supplement
% ???
There are several observations that can be made from generating likely pathways. One observation is that, while A1-A2-A3-A4 is a very stable 4-subunit subassembly, A1-A2-A3-A4-B5 is more stable than A1-A2-A3-A4-A5, the pentamer (see Figure~\ref{fig:stab-pent}). This indicates that the pentamer is not fully stabilized until the addition of the B subunit. Such reliance on dimeric interactions may correlate with the size specificity of the virus capsids, since such an interface will not be present in a T=1 shell. Another insight is that the A-C interface is not at all stable (the C-B or C-D interfaces are much better), and probably does not happen until much later in the assembly process or until other partners on hexameric interfaces provide necessary stabilization.

%\subsection{Likely Pathways}
% Try and implement the pathway generator, discuss and put values on all of this.

\subsection{Steady-State Concentration Calculations}
Transition pathways are useful for determining the most likely step in a single-step reaction. However, single-step reactions do not take into account one of the major driving forces for chemical reactions: concentrations the necessary reactants and products. If no reactants are available in a chemical reaction, it cannot take place; likewise, if the concentration of products is too high, the reaction will not proceed forward. For this reason, we also sought to provide a {\it global} view of the viral assembly in terms of concentrations of products and reactants. 

According to the \citeauthor{zlotnick05}\cite{zlotnick05}, the rate of formation of a subassembly consisting of $n$ copies of monomer $S$, $S^n$, is:
\begin{equation}
 \dv{\concnorm{S^n}}{t} = k_{\mathrm{assoc}}\concnorm{S^{n-1}}\concnorm{S}-k_{\mathrm{dissoc}}\concnorm{S^n},\label{eqn:dedt}
\end{equation}
where $\concnorm{S}$ is the concentration of a single subunit, and $\concnorm{S^{n-1}}$ is the concentration of $S^n$ without monomer $S$. At equilibrium ($\df{\concnorm{S^n}}/\df{t}=0$), the dissociation expression from Equation~\ref{eqn:dedt} indicates the concentration of free individual subunits that must remain:
\begin{equation}
 \frac{\concnorm{S^n}}{\concnorm{S^{n-1}}\concnorm{S}} = \frac{k_\mathrm{assoc}}{k_\mathrm{dissoc}}
\end{equation}

The ratio $k_\mathrm{assoc}/k_\mathrm{dissoc}$ at equilibrium is known as the {\it equilibrium constant}, $K_{S^n}$, and is directly related to the change in free energy, $\Delta G(S^n)$, as:
\begin{align}
 \Delta G(S^n) = -RT\ln K_{S^n} \\
 K_{S^n} = \exp{-\frac{\Delta G(S^n)}{RT}},
\end{align}
where $R$ is the gas constant (8.314 J mol$^{-1}$ K$^{-1}$) and $T$ is the temperature (298K).

Thus, if the concentration of the reactants, $\concnorm{S}$ and $\concnorm{S^{n-1}}$, and the change in free energy of the subassembly, $\Delta G(S^n)$, are known, then it is possible to compute the concentration of the product, $\concnorm{S^n}$:
\begin{equation}
 \concnorm{S^n} =  \concnorm{S^{n-1}}\concnorm{S}*\exp{-\frac{\Delta G(S^n)}{RT}}\label{eqn:concprod}
\end{equation}
This can be extended to subassemblies with generic reactants, such as $B5-C1$ and $A1$, and the product $A1-B5-C1$, as long as the concentrations of the reactants and $\Delta G$ of the product formation is known.

It should be noted here that for many chemical reactions, the rate of the reaction is determined by kinetics ($k_\mathrm{assoc}$ and $k_{\mathrm{dissoc}}$) and not by thermodynamics ($\Delta G$). However, when the values of $\Delta G$ are high enough, it can be assumed that 
the reaction will go to completion quickly, and the final ratio of products and reactants is equal to the equilibrium constant. As the $\Delta G$ values used in these experiments are very favorable (on the order of 300-3,000kJ/mol), this assumption was made through this section.

\subsubsection{Representing Capsid Assembly as a Graph MAP Problem}
Based on Equation~\ref{eqn:concprod}, it is easy to see that the concentration of a single product is dependent on the concentration of one or more reactants. If concentrations of subassemblies are represented by vertices in a graph, then dependencies can be represented by directional edges in this graph, that have weights proportional to the $\Delta G$ value for each formation. In this way, we can use a {\it graphical model} to describe the assembly process. If we initialize this graphical model with non-zero concentrations for the monomers (A, B, C, and D, for 1oph) and zero concentration for all other nodes and allow concentration to ``flow'' along edges from one node to another, then the {\it maximum a posteriori probability (MAP) estimate} is the steady-state of the graph, where no flow is happening. 

As the details of this representation are beyond the scope of this work, we will only state that we are using a Bayesian factor graph to represent the steady-state assembly and leave further details for the interested reader to the supplement of this paper. For figures in this paper, the Nudaurelia Capensis virus capsid was modeled with initial concentration of all monomers (A, B, C, and D) were typical concentrations at micromolar ranges (100e-nM). The message-passing algorithm was run until concentration change was below 1e-11M, summed over all concentration nodes.

%%% Results
\subsubsection{Steady-State Concentrations}
\begin{figure}
 \centering
 \includegraphics[width=\linewidth]{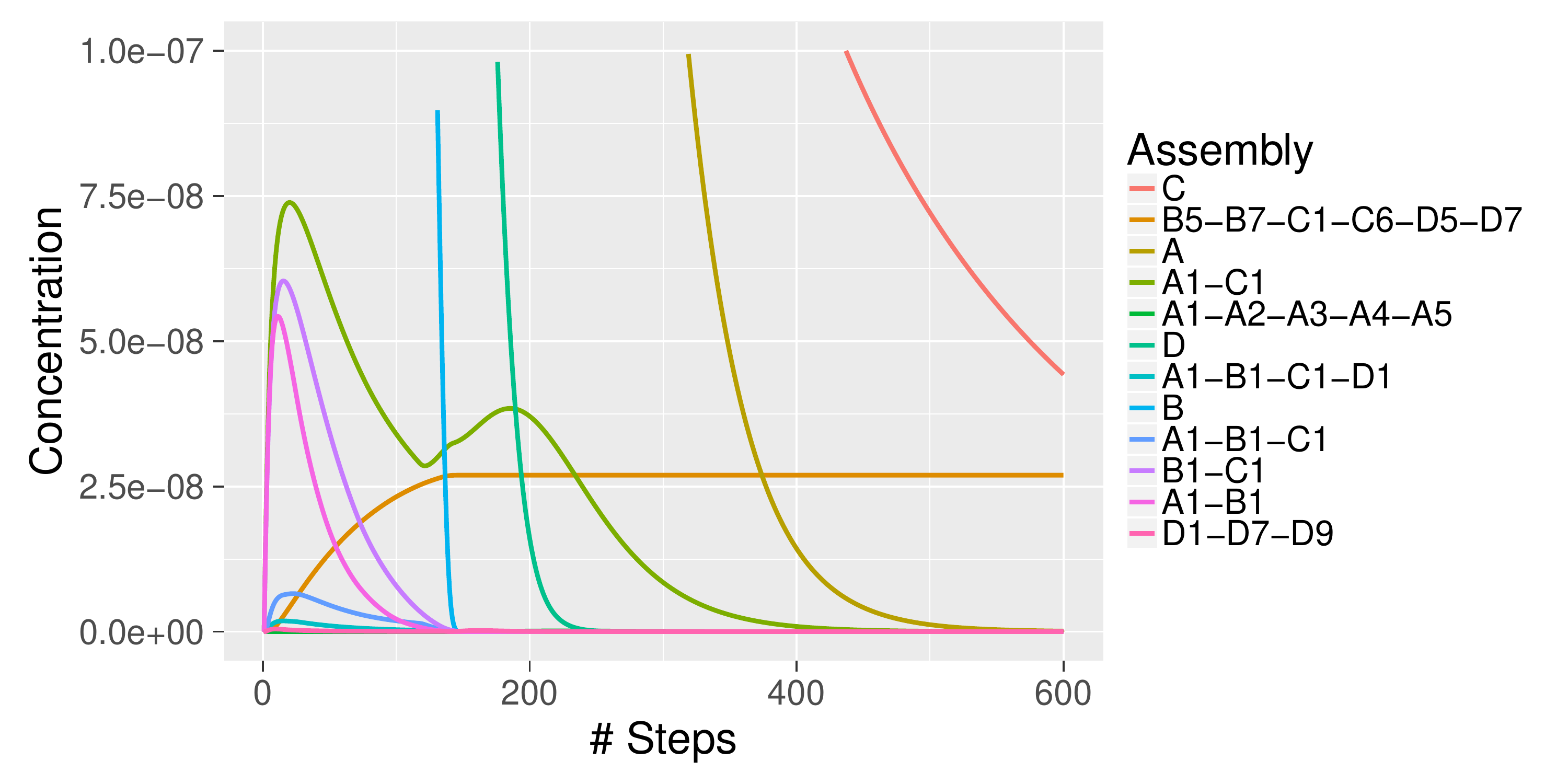}\\
 \caption{Change in concentration over time for several subassemblies of the Nudaurelia Capensis viral capsid. Legend is ranked according to concentration at step 600. The x-axis has been trimmed to emphasize initial concentration changes, as the steady-state concentrations were reached after 1300 steps of the algorithm.}\label{fig:steady-avg}
\end{figure}

% 1. Talk about differences in concentration for just PDB value vs sampling
%   - discuss relative concentrations of pentamer, trimer, etc
%           - mention this is partial, after only 6-mers. might have seen different results after running with more subassemblies
Figure~\ref{fig:steady-avg} shows the distribution over successive steps of the message passing algorithm, which attempts to model the self-assembly of the capsid. For the most part, the assemblies with higher concentrations at the steady state (final step of the algorithm) are those with more subunits (e.g., the concentration of the hexamer B5-B7-C1-C6-D5-D7 was 26nM, several orders of magnitude higher than the other subassemblies). This observation would suggest that the subassembly formation largely proceeded toward completion of products, and that the limiting factor was concentration of the products, as was expected. 

An important point to note is that this graph would be different if subassemblies of size greater than 6 were also included, as the intermediate products would be quickly consumed. From Figure~\ref{fig:steady-avg}, this phenomenon can already be observed; for example, the intermediate product A1-C1 has an initial high concentration, but then quickly drops off as it is used for later products, such as A1-B1-C1-D1. This also explains why the concentration of monomer C decays so slowly, as there are fewer beneficial reactions involving C (see, for e.g. Figure~\ref{fig:tri-wilcox}, where the products involving C are not highly ranked). This might suggest that the configuration of C with the rest of the capsid is meant as a stabilizing subassembly, and is not used until much later in the assembly process.
%   - then discuss differences between values from sample vs values from PDB

\begin{figure}
 \centering
 \includegraphics[width=\linewidth]{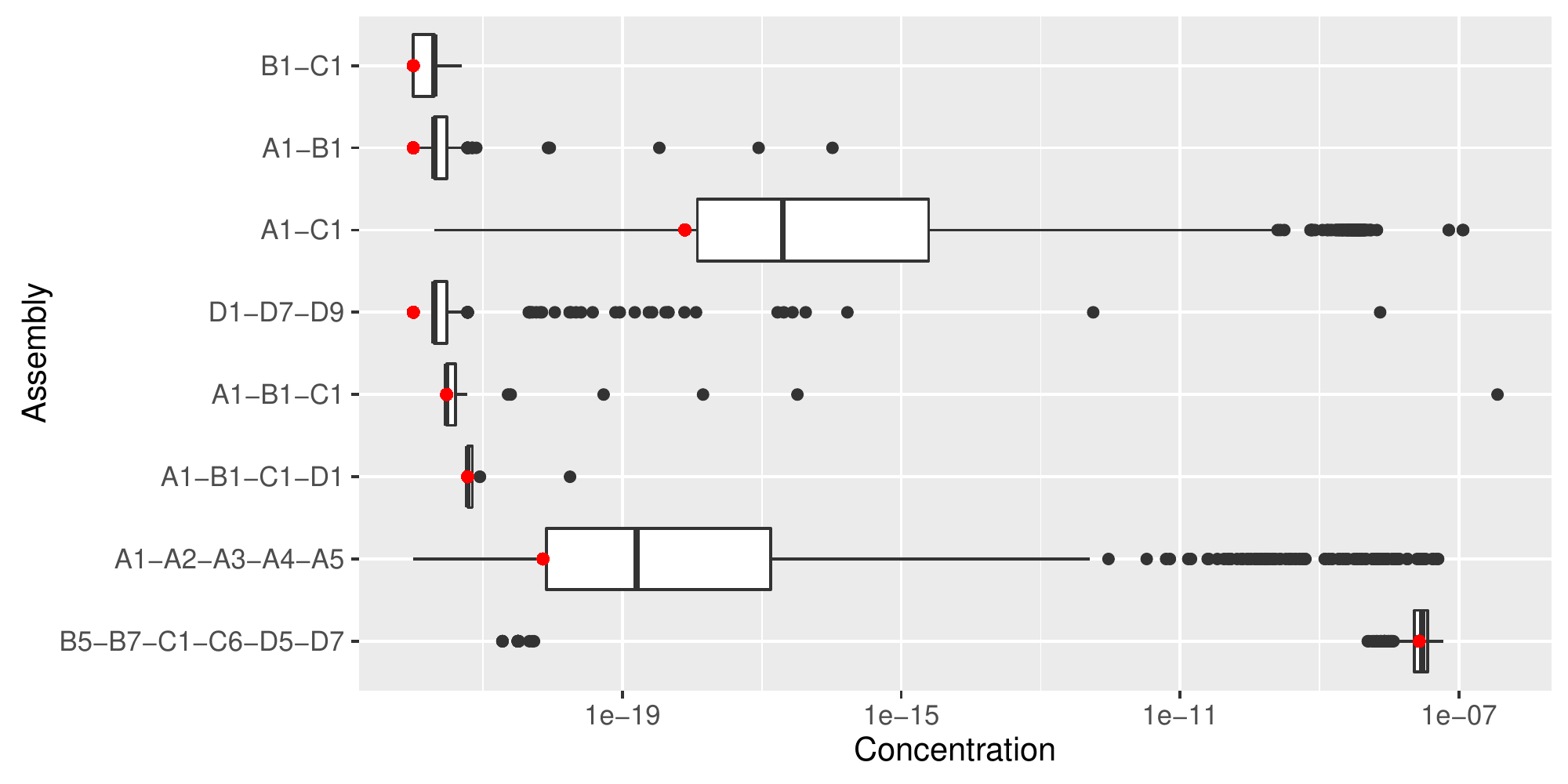}
 \caption{Distribution of concentrations over different input $\Delta G$ values, for various subassemblies of the Nudaurelia Capensis viral capsid. X-axis is log-scale to emphasize differences. Red point is value computed when using just the original PDB.}\label{fig:steady-distr}
\end{figure}

% 2. Talk about initial concentrations, and what is getting created / consumed first, etc
%   - discuss peaks, and what they mean (getting consumed)
Finally, we can also plot the distribution of all possible steady state concentrations, shown in Figure~\ref{fig:steady-distr}. For this plot, initial values of $\Delta G$ were taken from the {\it distribution} of possible values for each subassembly, and then the steady-state assembly algorithm was run as usual. Distribution of final concentration of subassemblies were plotted as box-and-whisker plots. The red dot shows the value computed when using $\Delta G$ computed on the unperturbed PDB structure. This plot shows several things: first, that the distributions of final concentrations differ greatly across subassemblies, and second, that the value computed on the original PDB does not represent the true average across all samples. 

\section{Conclusions}
Most of existing research in assembly pathway prediction/analysis of virus capsids have relied on the final configuration of the capsid to determine the configuration of the intermediate states. This assumption is overly simplified since the capsid proteins may undergo conformational changes, binding interfaces may change to allow binding with another subassembly, etc.\ throughout the assembly process until stabilization. To better capture this phenomenon, we have developed a statistical-ensemble based approach which sufficiently samples the configurational space of each monomer and the relative local orientation between monomers to capture the uncertainties in their binding. Essentially, instead of modeling each subassembly as a static configuration, they are modeled as distributions of possible configurations. This allows us to compute the free energy of a subassembly, and the binding free energy on a possible assembly edge as distributions so that statistical guarantees of accuracy can be additionally derived for each of the resulting assemblies. 

Unlike traditional approaches where pentamers+hexamers, or trimers are used as fundamental building blocks in the assembly pathway analysis, we use individual monomers as our starting constituents, and consider all possible unique subassemblies (modulo symmetry), of sizes up to 6. The primary aim is to quantitatively understand the formation of the larger building blocks (i.e. trimers, pentamers and hexamers). 

We additionally, adapted the Wilcoxon measure to provide a way to compare the distributions and determine the most likely subassemblies that can be generated in any step. Using this score as weights on an assembly graph revealed that there are some low-energy subassemblies that are unlikely to be formed because there are poor transitions along the path that forms said subassembly. Finally, we extended the assembly prediction to factor concentrations of the constituents. The assembly graph was converted to a Bayesian factor graph where the final concentrations of the subassemblies are posed as a graphical maximum a posteriori problem. Transition probabilities were set up based on the equilibrium constant computed from the binding free energy, and both forward (association), and backward (dissociation) reactions were allowed. The result showed expected patterns, e.g. dimers A1-B1, A1-A2 etc. getting produced at a fast rate and then being consumed as other subassemblies become available, forming the larger subassemblies A1-B1-C1 (trimer), A1-A2-A3-A4-A5 (pentamer), etc. As the concentrations reach their steady state, larger particles had higher final concentrations, as was expected.

In summary, we contend that the use of {\it ensemble distributions of molecules}, instead of single conformations, allows one to make statistical inferences about the stability of molecular subassemblies. We have shown that, a full distribution of possible subassemblies  is not obtainable  if one was to use assembly combinations only from the original PDB conformation.  This could often lead to erroneous conclusions. Use of a  statistically rigorous procedure such as the one advocated in this paper, yields inferences on capsid assembly  can be made with  confidence.

\section{Supporting Information Available}
We have added several files containing more in depth description of our methods, additional figures, and some data as supporting information. A list of the files and there content follows:
\begin{itemize}
	\item Supporting Information for Publication: Contains details of the Bayesian Factor Graph construction, and additional figures S1-S10.
	\item {\tt graphical\_model\_edges.avg.txt}: The graphical model for MAP analysis in ASCII format.
	\item {\tt Wilcoxon\_graph.txt}: The complete graph corresponding to Figure \ref{fig:transition-graphs}.
\end{itemize}

\section{Acknowledgements}
This research was supported in part by NIH grant R01-GM117594-0 and Sandia subcontract SNL-1439100.

%%%%%%%%%%%%%%%%%%%%%%%%%%%
% Appendix
%%%%%%%%%%%%%%%%%%%%%%%%%%%
\beginsupplement
\section{Supplemental Information}
%\subsection{Calculating the MAP Estimate Using Message Passing}
\subsection{Representing Viral Assembly as Bayesian Factor Graph}
\begin{figure}
\centering
 \includegraphics[width=0.38\linewidth]{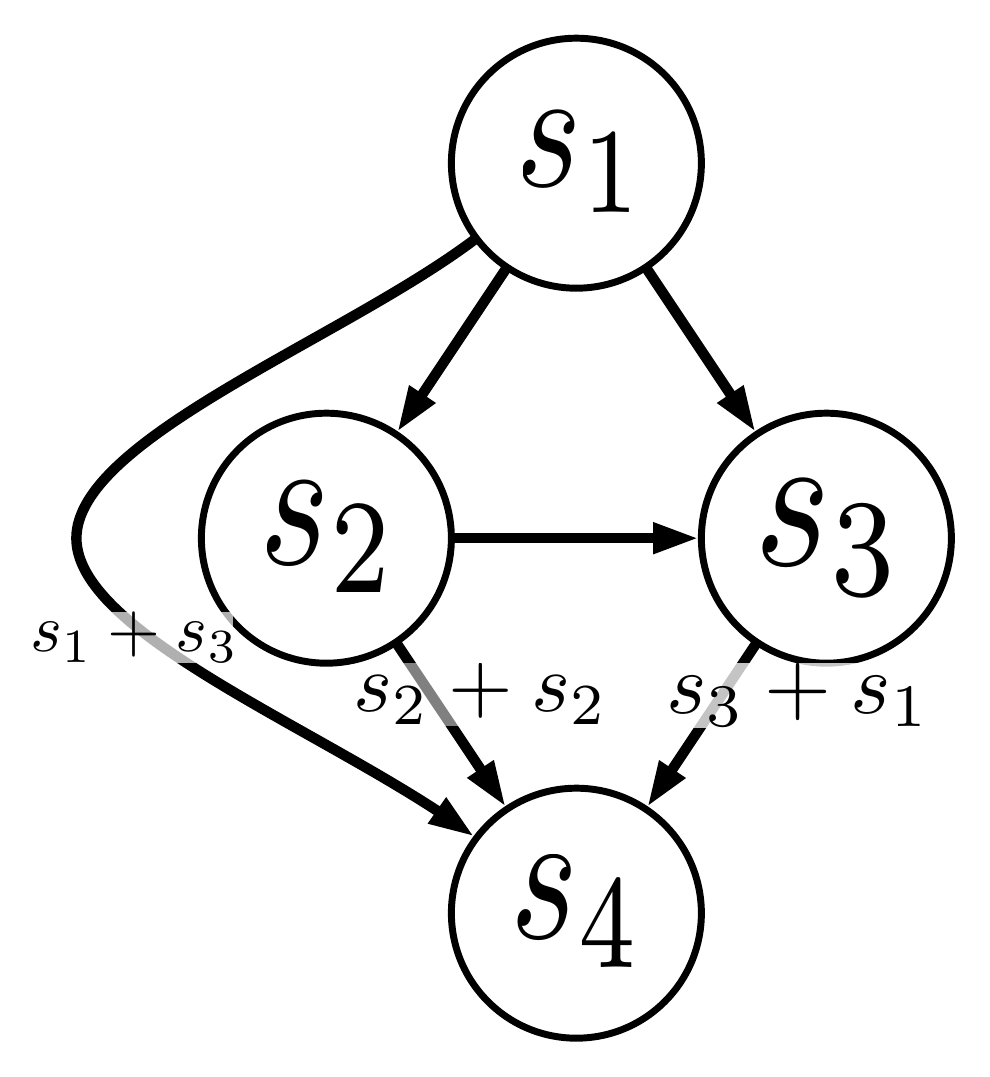}\quad
 \includegraphics[width=0.48\linewidth]{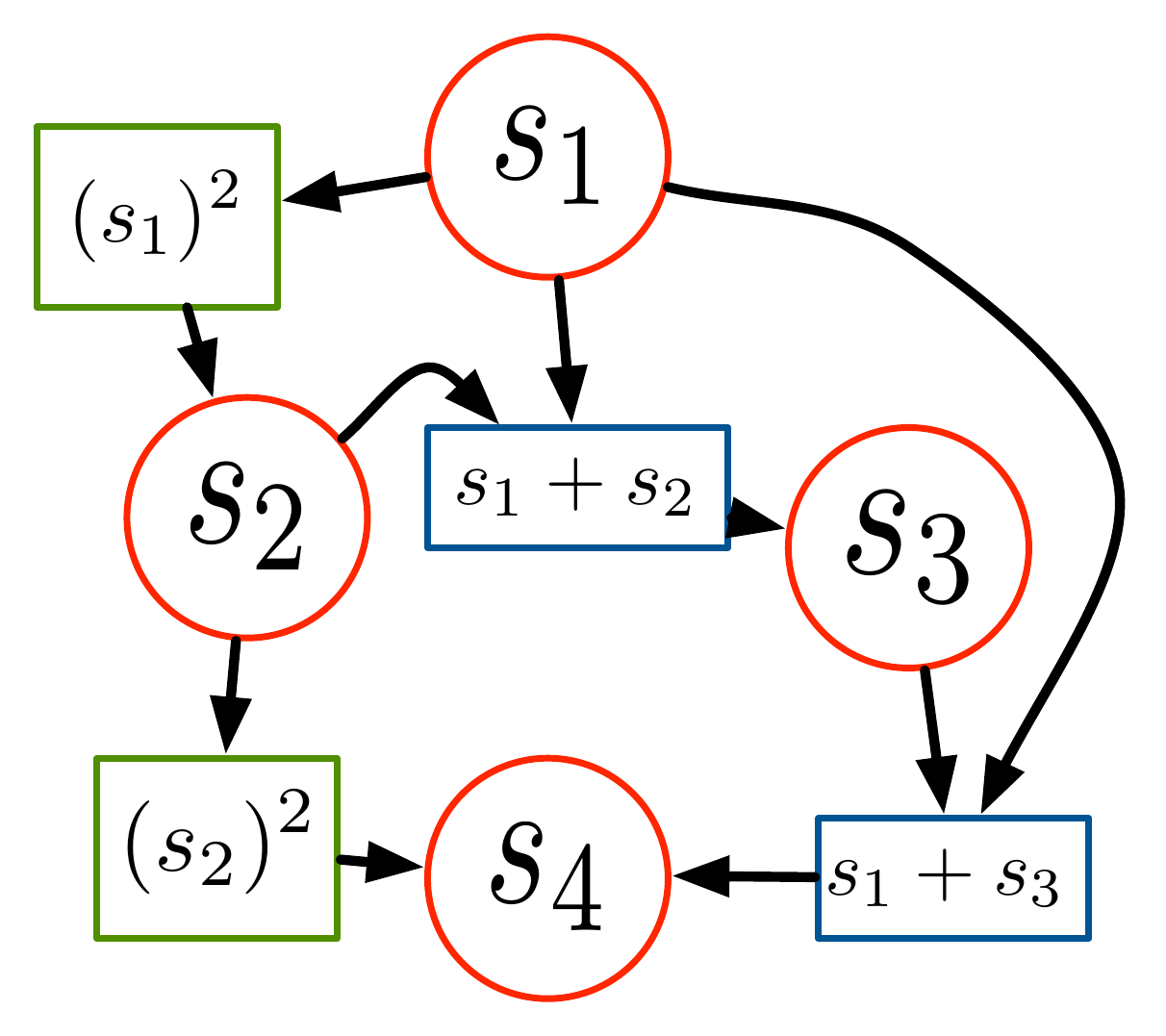}
 \caption{Two different graphical representations of virus formation. Left: a simple directed Bayesian network; right: the same network represented by a factor graph, where {\it factor} nodes are rectangles, representing the combination of a set of smaller nodes.}\label{fig:graphs}
\end{figure}

For a simple virus (T=1) consisting of $n$ identical subunits, it is easy to represent the formation of a virus as a Bayesian network consisting of $m$ nodes, where each node represents a possible subassembly, $s_k$, and the transition probabilities are the rates of formation. A node, $v_{s_k}$, representing $s_k$, would have incoming edges from all $s_{i}$ and $s_j$, where $i+j=k$ (e.g. $s_{k-1}$ and $s_1$, as well as $s_{k-2}$ and $s_2$, etc.). See Figure~\ref{fig:graphs} (left), where a larger subassembly, $s_4$, is created from the combination of $s_1+s_3$ or $s_2+s_2$. Edges leading into $s_4$ are labeled with the specific combination. The concentration $\concnorm{s_4}$ is dependent on $\concnorm{s_3}$, $\concnorm{s_2}$, and $\concnorm{s_1}$, as well as the corresponding $\Delta G$ values.

\paragraph{Bayesian Factor Graph}
One of the limitations with a traditional Bayesian network is that a simple edge weight does not contain all the information necessary to determine forward and backward formation. The dependencies of $s_4$, for example, are $s_1$, $s_2$, and $s_3$. However, $s_4$ is constructed of two $s_2$ subunits, but only one $s_1$ subunit. Instead of a Bayesian network, we can represent the viral assembly process as a Bayesian factor graph, where the {\it factor} nodes are states that contain this additional information. Thus, the only dependencies for $s_4$ are $s_2+s_2$ or $s_1+s_3$, not the individual subunits. Because we are modeling the creation of a virus from the addition of single subunits, we will enforce an additional constraint that each factor node must have exactly two incoming edges and exactly one outgoing edge.

\subsection{Solving MRFs Using Message Passing}
The traditional method for solving MRFs is through a technique called {\it belief propagation} or {\it message passing}, where each node in the graph with ``propagate'' its ``belief'' about the current state of the network to its neighbors\cite{yedidia2003}. Traditional belief propagation is performed by the sum-product method\cite{kschischang2001}, which will be adapted for our use here.

Let $r$ be a subunit that is used to produce more than one product, e.g. $r$ and $r_1$ form $p_1$, and $r$ and $r_2$ form $p_2$. Then the ratio of the two products can be determined from Equation~\ref{eqn:concprod}. If we assume that the concentration of $r_1$ and $r_2$ are equal, then:
\begin{align}
 \frac{\concnorm{p_1}}{\concnorm{p_2}} & = \frac{\exp{-\frac{\Delta G(p_1)}{RT}}}{\exp{-\frac{\Delta G(p_2)}{RT}}} \\
   & = \exp{\frac{1}{RT}\left(\Delta G(p_2)-\Delta G(p_1)\right)}\label{eqn:concratio}
\end{align}
If we let $e(p_2)$ be the amount in the exponent, i.e.:
\begin{align}
 e(p_2) = \exp{-\frac{\Delta G(p_2)}{RT}},
\end{align}
then we can rewrite Equation~\ref{eqn:concratio} as:
\begin{equation}
 \frac{\concnorm{p_1}}{\concnorm{p_2}} = \frac{e(p_1)}{e(p_2)}\label{eqn:concratsimp}
\end{equation}

For a set of $k$ potential products, $p_1\ldots p_k$, the proportional concentration of reactant $r$ that will be used to form product $p_i$, $\lambda(p_i)$, can be written from Equation~\ref{eqn:concratio} and \ref{eqn:concratsimp} as:
\begin{equation}
 \lambda(p_i) = \frac{e(p_i)}{\sum_{j=1}^k e(p_j)} \label{eqn:propconc}
\end{equation}

For the sake of notation, we will also define the reverse exponent amount for reactant $r$, $e^{-1}(r)$ as
\begin{equation}
 e^{-1}(r) = \exp{\frac{\Delta G(r)}{RT}}
\end{equation}
and the reverse proportion of a reactant, $\lambda^{-1}(r_j)$, over a set of potential reactants, $r_1\ldots r_k$, as:
\begin{equation}
 \lambda^{-1}(r_j) = \frac{e^{-1}(r_i)}{\sum_{i=1}^k e^{-1}(r_i)} \label{eqn:propconc-rev}
\end{equation}

\begin{figure}
 \centering
 \includegraphics[width=3in]{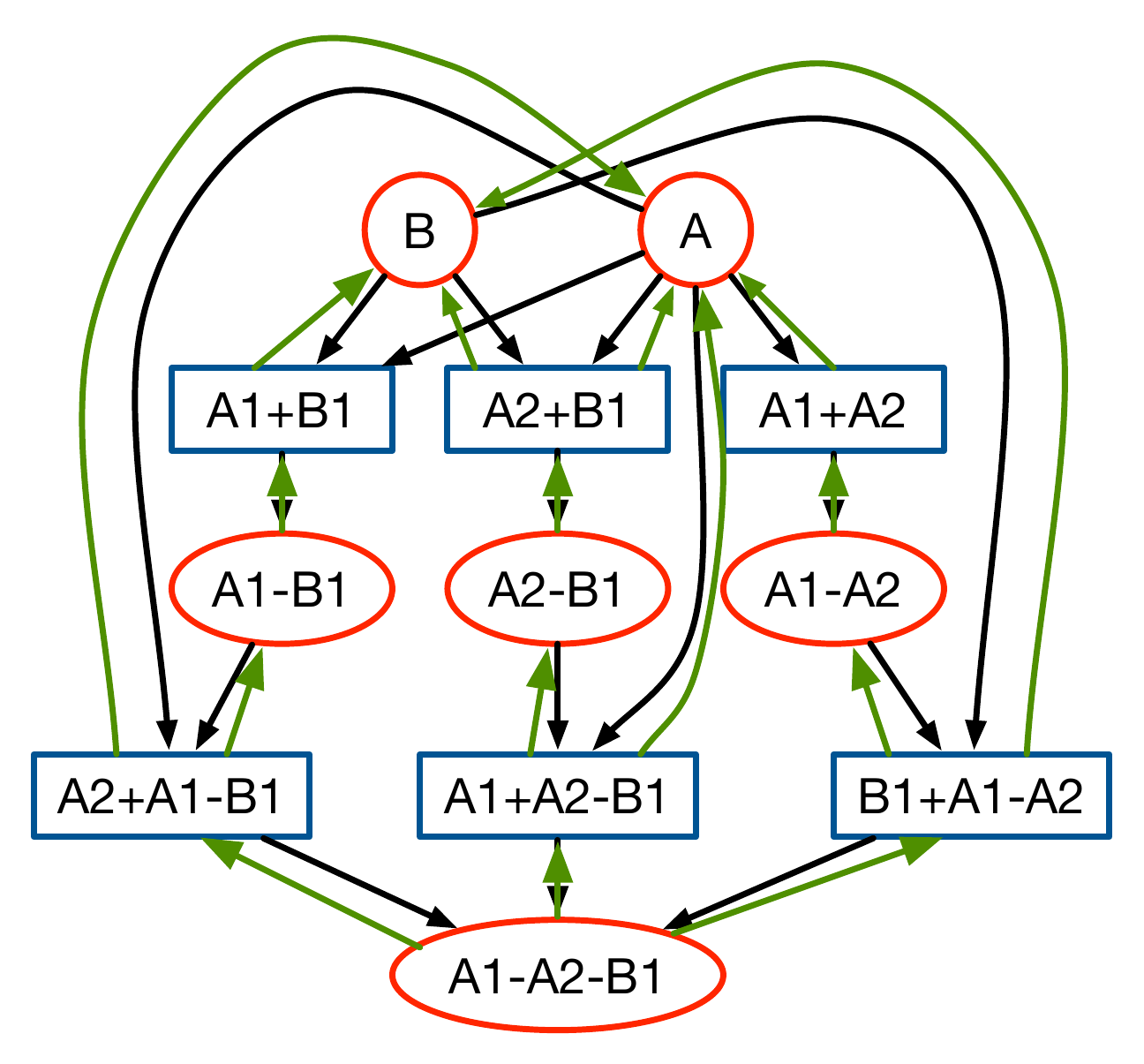}
 \caption{Factor graph for the formation of $A1-A2-B1$ from all possible units. Variable nodes are red circles, factor nodes are blue rectangles, forward edges are black, and backward edges are green.}\label{fig:a1a2b1}
\end{figure}

Let $G=[E,V,F]$ be a bipartite graph, $G$, with edges $E$ and nodes divided into two disjoint groups $V$ (variable nodes, or concentrations of subassemblies) and $F$ (factor ``pool'' nodes, or hidden nodes where concentrations `pool' before being assembled). Then the only messages that are passed are from $v\in V$ to $a\in F$ and vice versa, and not between two members of the same set. We will also define two different kinds of edges, {\it forward}, and {\it backward}. Forward edges represent the formation of a product by two reactants, and backward edges the break-down of products into reactants. See Figure~\ref{fig:a1a2b1} for an example of the formation of $A1-A2-B1$ from all possible reactants. The we will define four types of messages here:

Messages along the forward edges are:
\begin{itemize}
  \item $\mu_{v\rightarrow a}(\concnorm{v})$: A message along from a variable (concentration) node to a factor (pool) node describing the belief of $\concnorm{v}$, that will be available at pool node $a$. Pool node $a$ will be used to pool the subunits for subassembly $v+x^*$, where $x^*$ is some (possibly different) subunit.
    \begin{equation}
      \mu_{v\rightarrow a}(\concnorm{v}) = \lambda(v-x^*) * \concnorm{v},
    \end{equation}
    where $\lambda(\concnorm{v-x^*}$ is defined according to Equation~\ref{eqn:propconc}, the predicted amount of $v$ that will be available at pool node $a$.
  \item $\mu_{a\rightarrow v}(\concnorm{v})$: A message from factor (pool) node $a$ to variable (concentration) node $v$ will be the amount of $v$ that is created. Since each factor node has exactly two incoming edges, let us call these edges $a[1]$ and $a[2]$. Then the message passed to $v$ will be:
    \begin{equation}
      \mu_{a\rightarrow v}(\concnorm{v}) = a[1]a[2]*e(v)
    \end{equation}
\end{itemize}

Messages along the backward edges are:
\begin{itemize}
 \item $\mu_{v_{a[j]}\leftarrow a}\left(\concnorm{v_{a[j]}}\right)$: Because the forward message from a reactants pool node to the products concentration node will not always consume all of the products, it becomes necessary to send a message back to the reactants containing the concentration amount not used. This message will be:
 \begin{equation}
  \mu_{v_{a[j]}\leftarrow a}\left(\concnorm{v_{a[j]}}\right) = a[j] - \min\left(a[1], a[2]\right),\quad j=\{1,2\}
 \end{equation}
 \item $\mu_{a_i \leftarrow v}(\concnorm{a_i})$: The concentration at product $v$ will come from several sources; the total product must therefore reach equilibrium with all of them. Thus, the message back from the product node $v$ to the factor node $a$ about product $a[1]$ or $a[2]$ will be:
 \begin{equation}
  \mu_{a[1] \leftarrow v}\left(\concnorm{a_1}\right) = \mu_{a[2] \leftarrow v}\left(\concnorm{a_2}\right) = \lambda^{-1}(v) * \concnorm{v},
 \end{equation}
 where $\lambda^{-1}(v) * \concnorm{v}$ is defined according to Equation~\ref{eqn:propconc-rev}.

\end{itemize}

\paragraph{Message Passing Algorithm}
The message passing algorithm will be run in the following steps. Let $T_0$ be the initial concentration of monomer subunits, and let $k$ be the size (number of monomers) of the largest subassembly.
\begin{enumerate}
  \item Initialize monomers to starting concentration, $T_0$
  \item Until no changes in concentrations have been made:
    \begin{enumerate}
    \item For each size $s=1$ to $s=6$ (monomer to largest):
      \begin{enumerate}
      \item pass messages along forward edges from $v$ of size $s$ to $f$ of size $s+1$
      \item pass messages along forward edges from $f$ of size $s+1$ to $v$ of size $s+1$
      \end{enumerate}
    \item For each variable node, $v_i$, set the concentration $\concnorm{v_i}$ to the sum of all messages along forward edges, $\mu_{*\rightarrow v_i}(\concnorm{v_i})$
    \item For each size $s=6$ to $s=1$ (largest to monomer):
      \begin{enumerate}
      \item perform message along all {\it backward} edges from $v$ of size $s$ to $f$ of size $s$
      \item perform message passing along {\it backward} edges from all $f$ of size $s$ to $v$ of size $s-1$
      \end{enumerate}
    \item For each variable node, $v_i$, add $\mu^*_{a\rightarrow v_i}(\concnorm{v_i})$ to the current concentration, $\concnorm{v_i}$
    \end{enumerate}
  \item Report the concentration of all $v$ nodes

\end{enumerate}

\clearpage
%\newpage
\subsection{Supplemental Figures}
\begin{figure}[h!]
\includegraphics[width=0.48\linewidth]{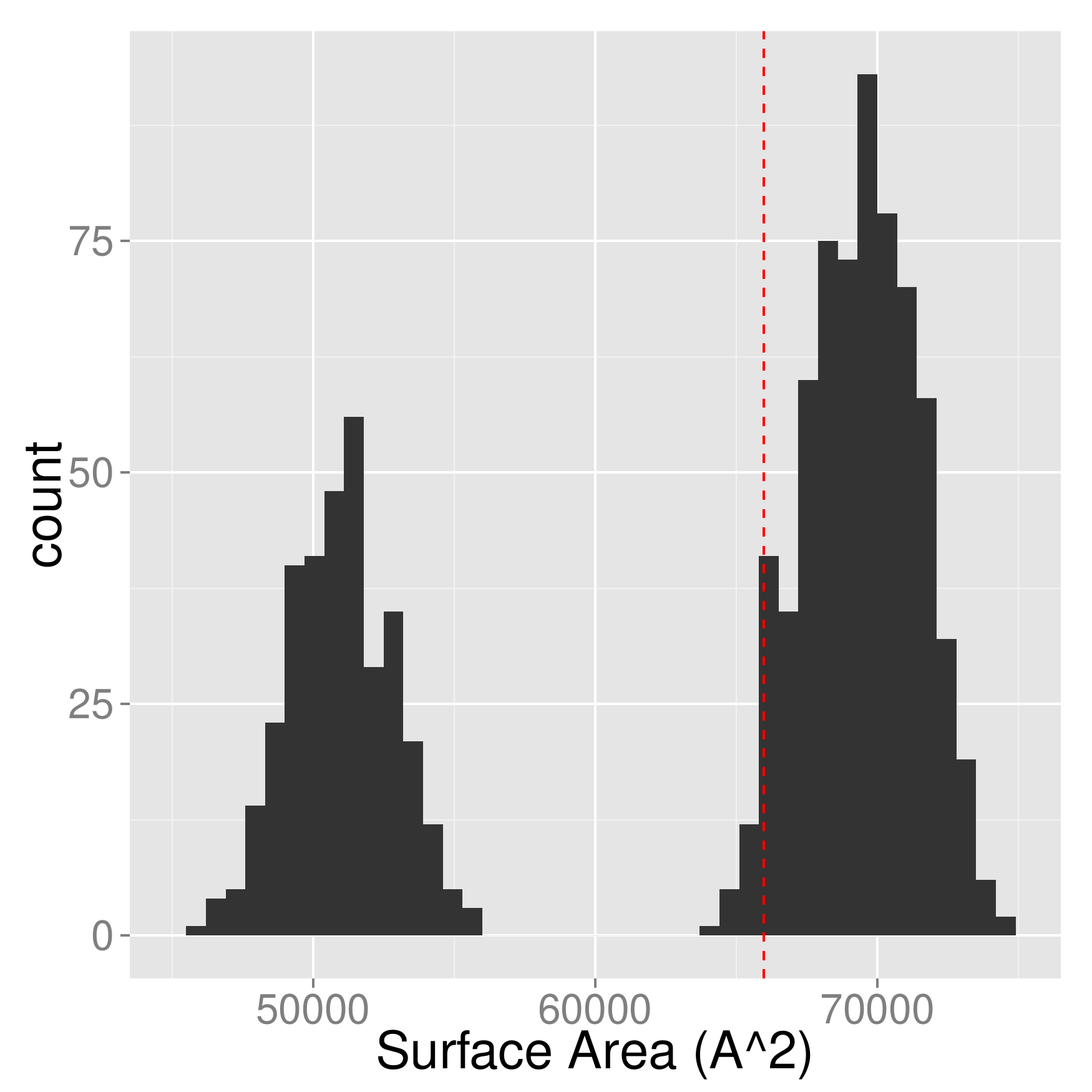}
\includegraphics[width=0.48\linewidth]{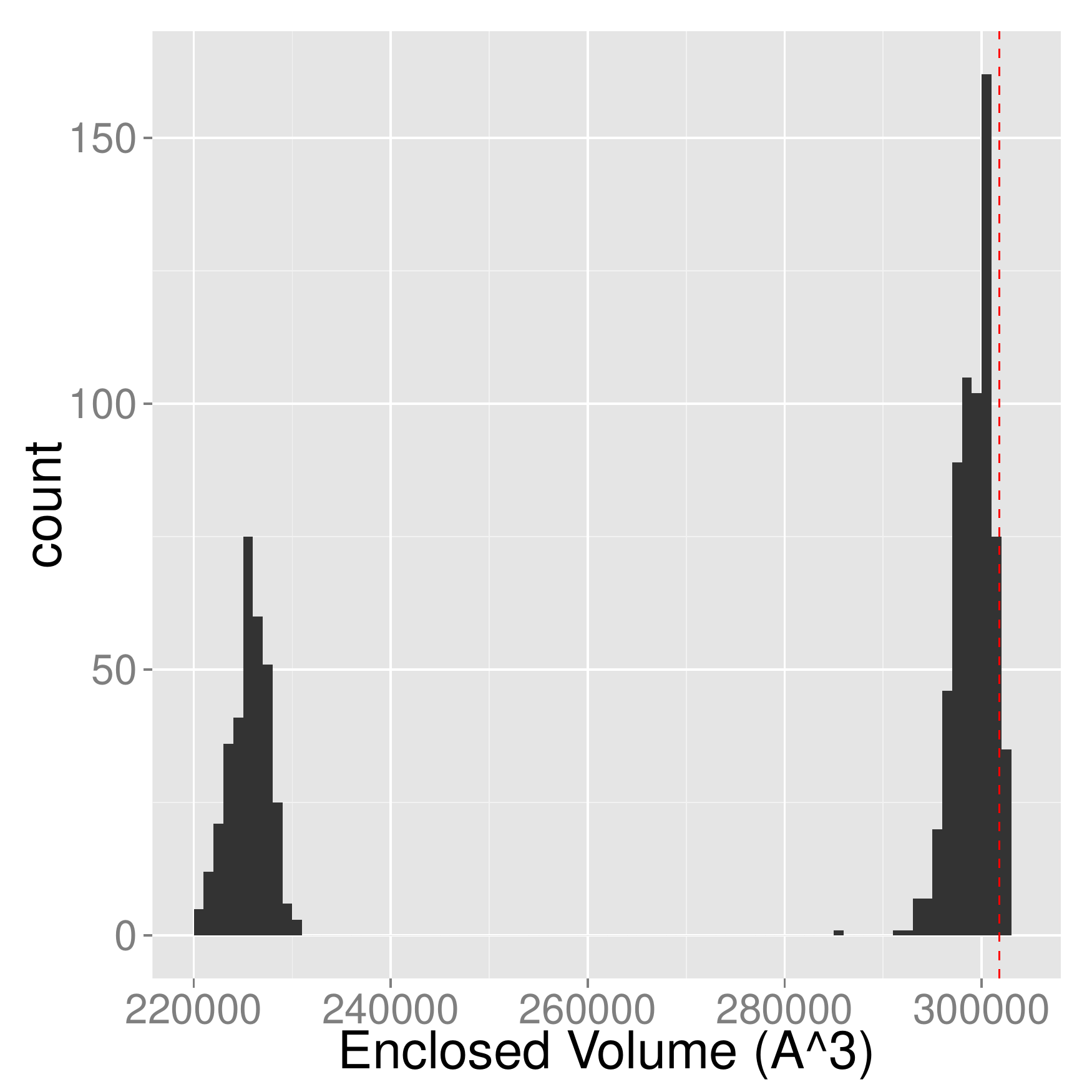}\\
\includegraphics[width=0.48\linewidth]{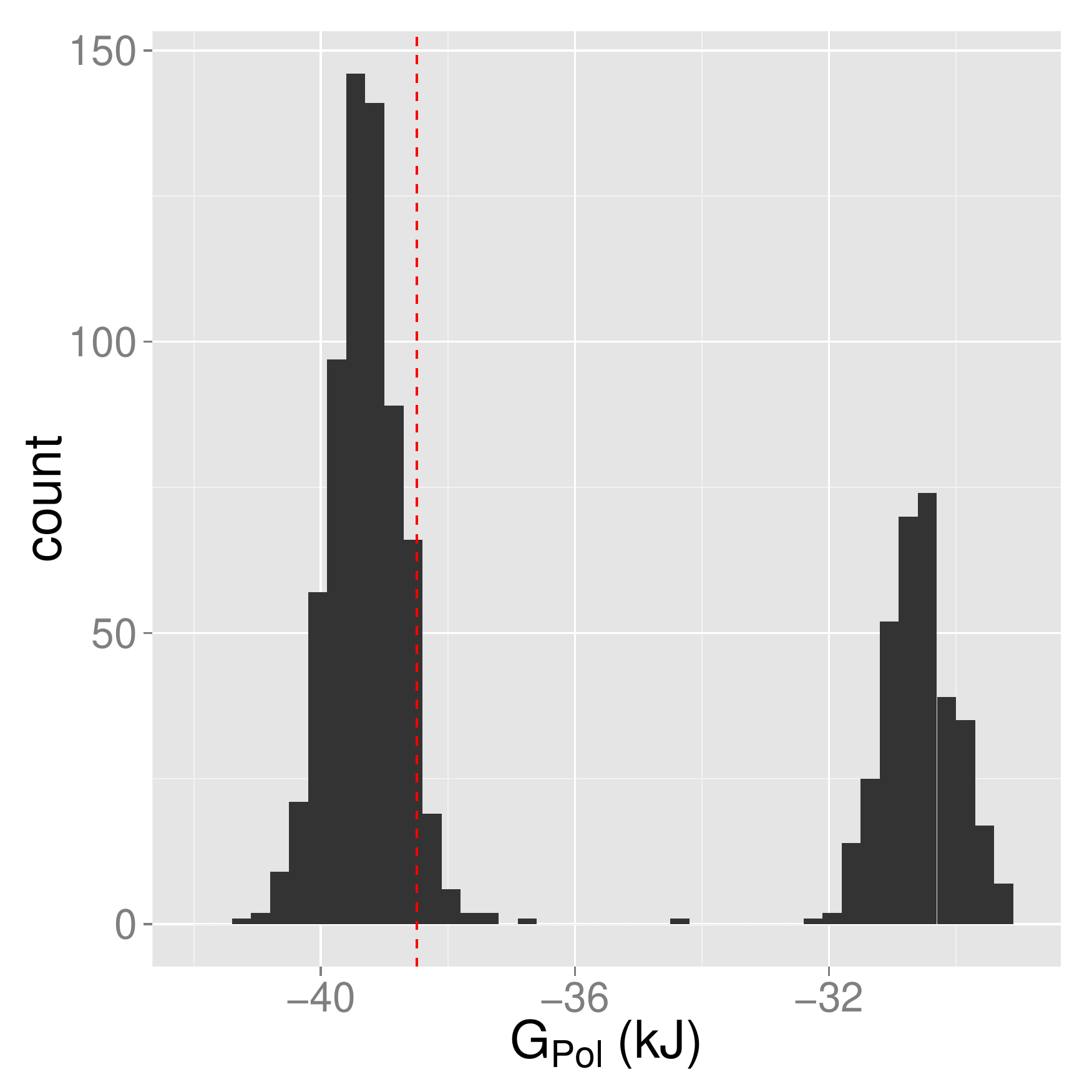}
\includegraphics[width=0.48\linewidth]{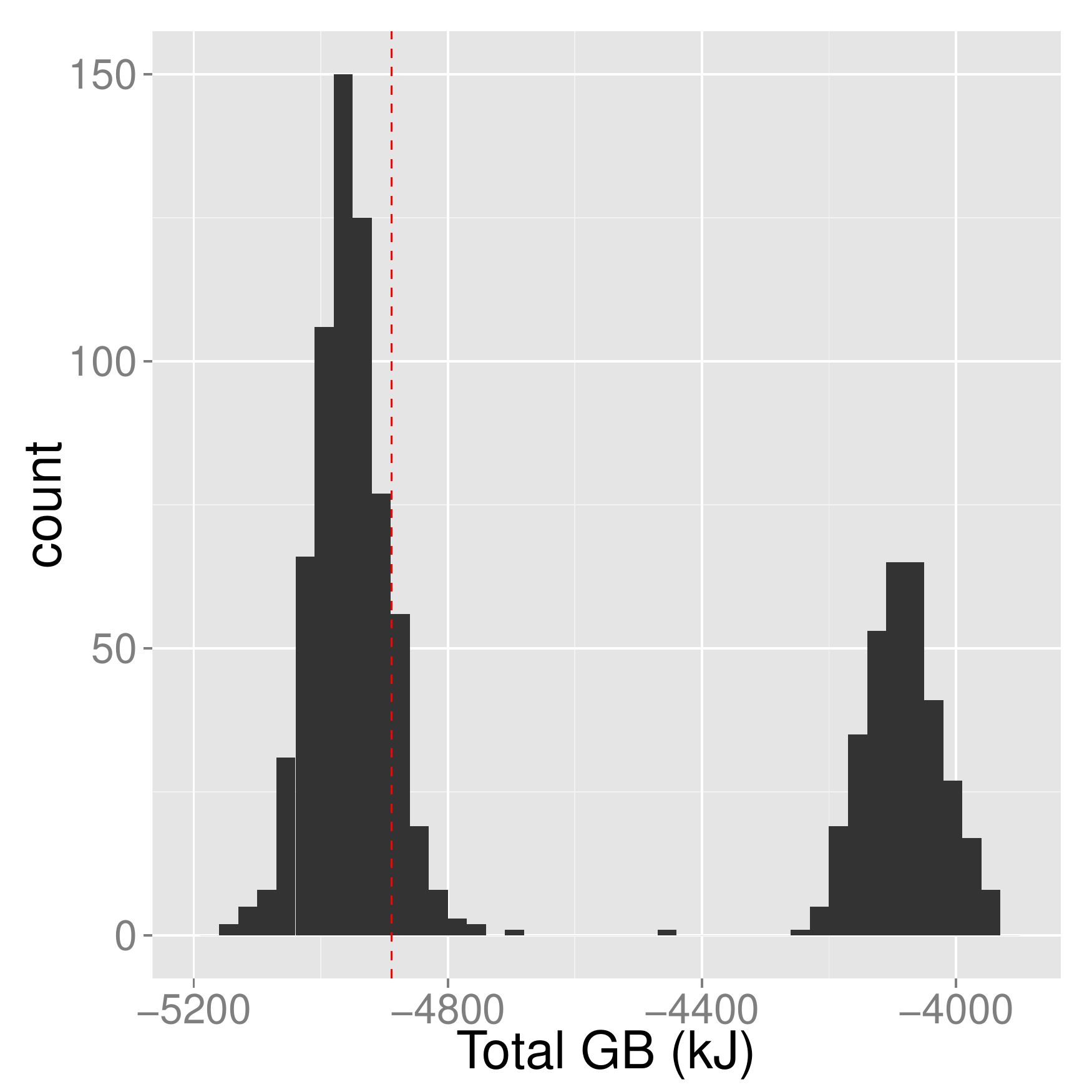}
\caption{Histogram plots of exposed surface area, enclosed volume, $G_{pol}$ and total energy (GBSA model) for all samples of B1-B5-B7-C1. Dotted vertical lines show the value computed on the un-sampled molecule. Bimodal nature of plots reveal low surface area between B7 and the rest of the complex. See the surface representation in \ref{fig:s-surfs}.}\label{fig:s-b1-b5-b7-c1}
\end{figure}

\begin{figure}
\includegraphics[width=0.48\linewidth]{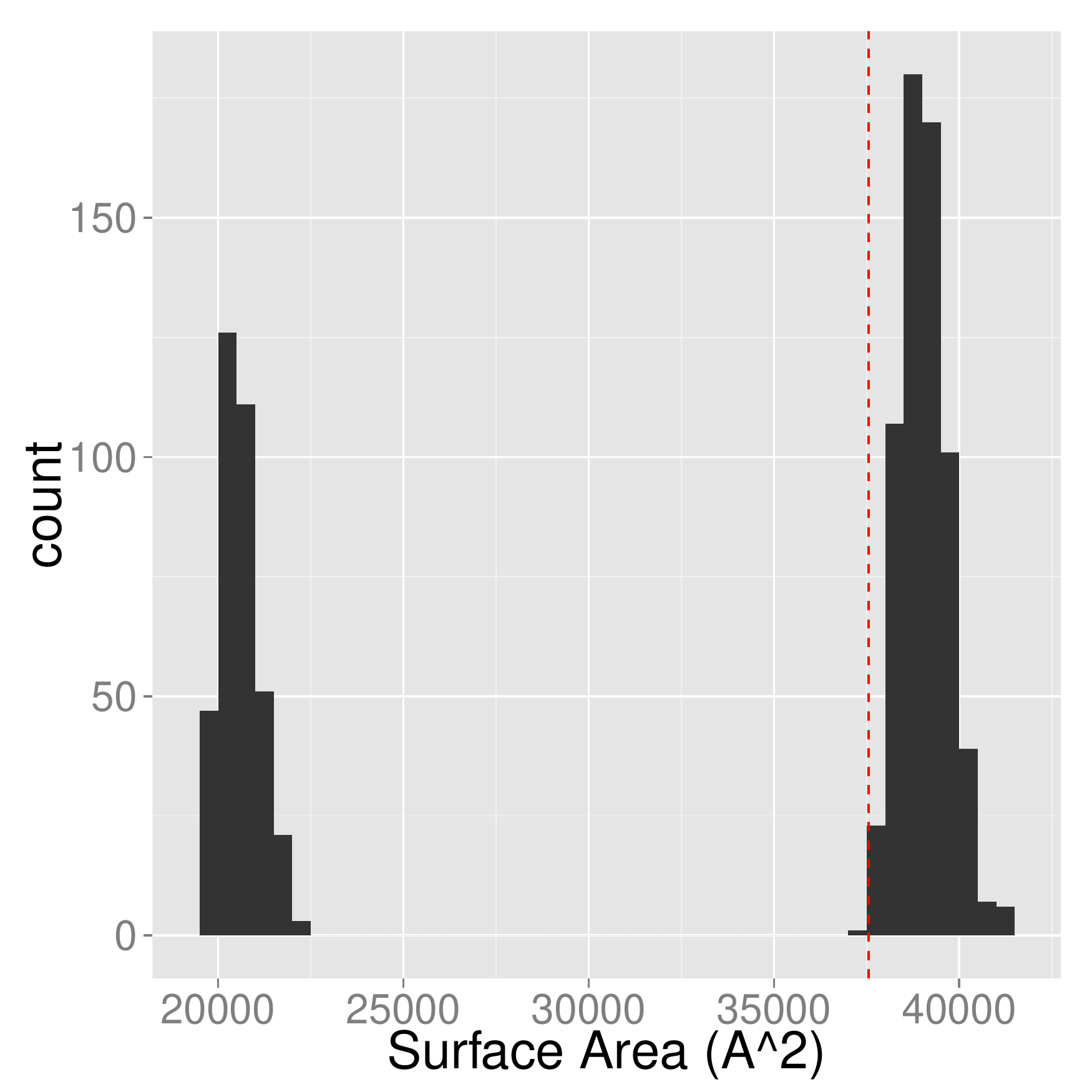}
\includegraphics[width=0.48\linewidth]{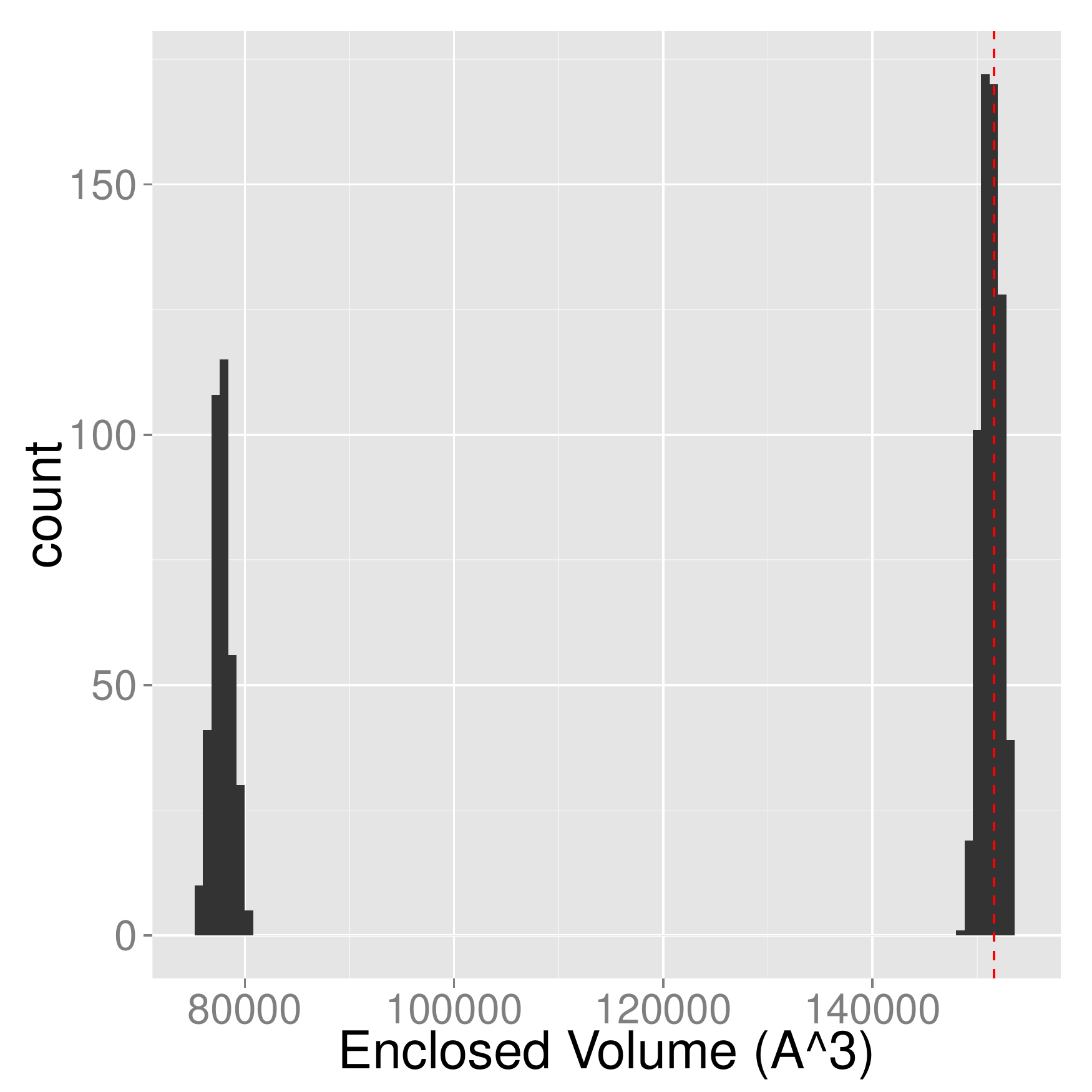}\\
\includegraphics[width=0.48\linewidth]{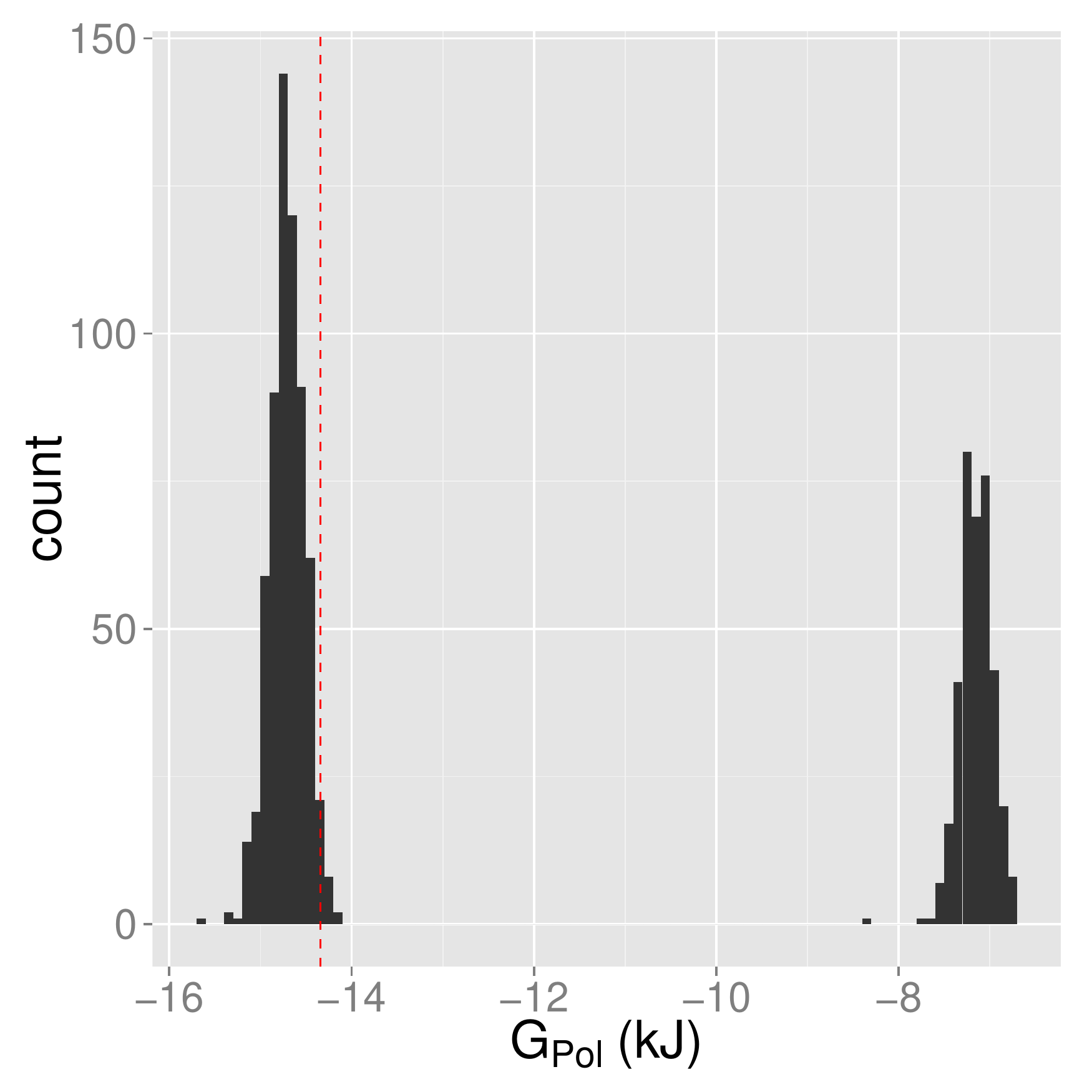}
\includegraphics[width=0.48\linewidth]{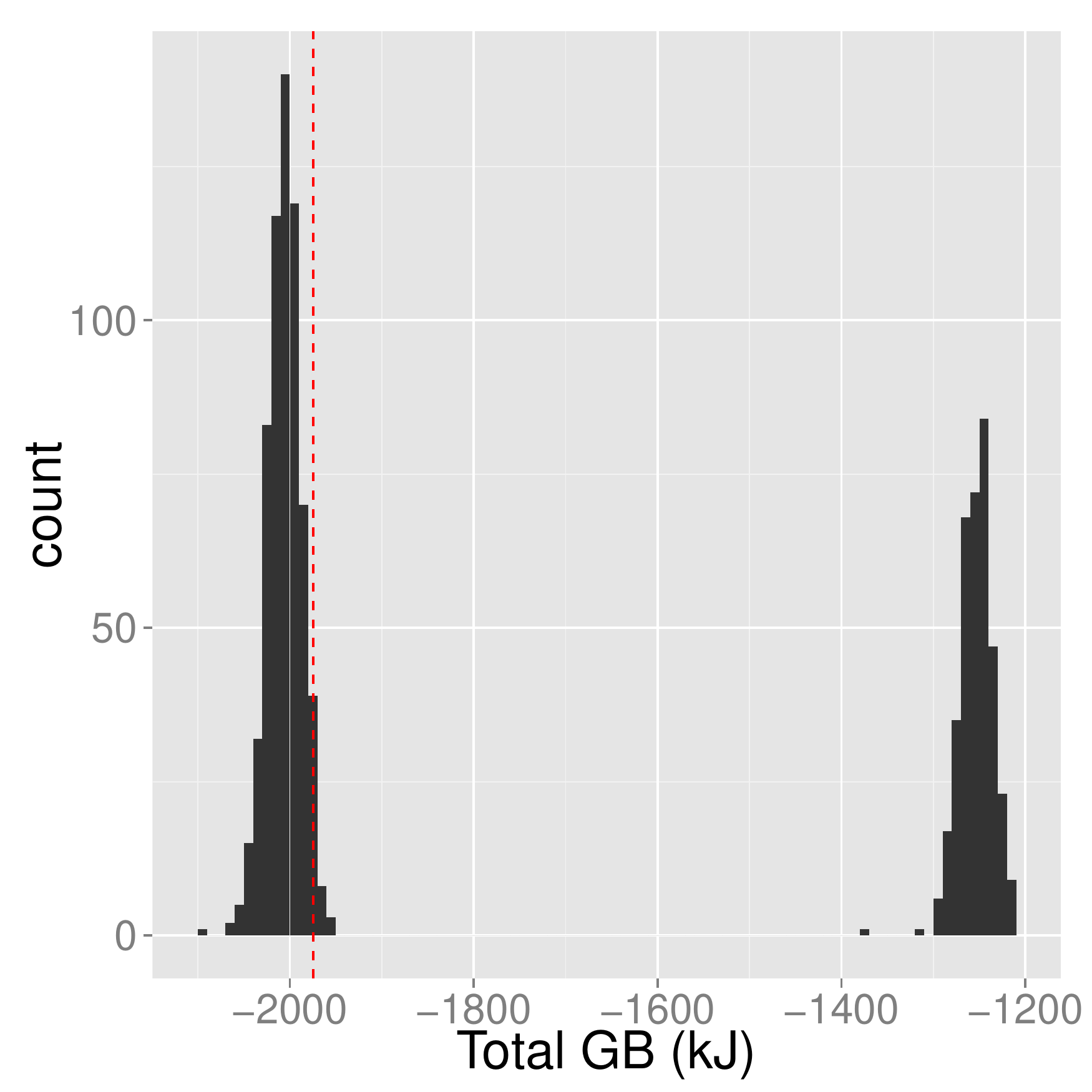}
\caption{Histogram plots of exposed surface area, enclosed volume, $G_{pol}$ and total energy (GBSA model) for all samples of B5-D7. Dotted vertical lines show the value computed on the un-sampled molecule. Bimodal nature of plots reveal low surface area between B5 and D7. See the surface representation in Figure \ref{fig:s-surfs}.}\label{fig:s-b5-d7}

\end{figure}
\begin{figure}
 \includegraphics[width=0.3\linewidth]{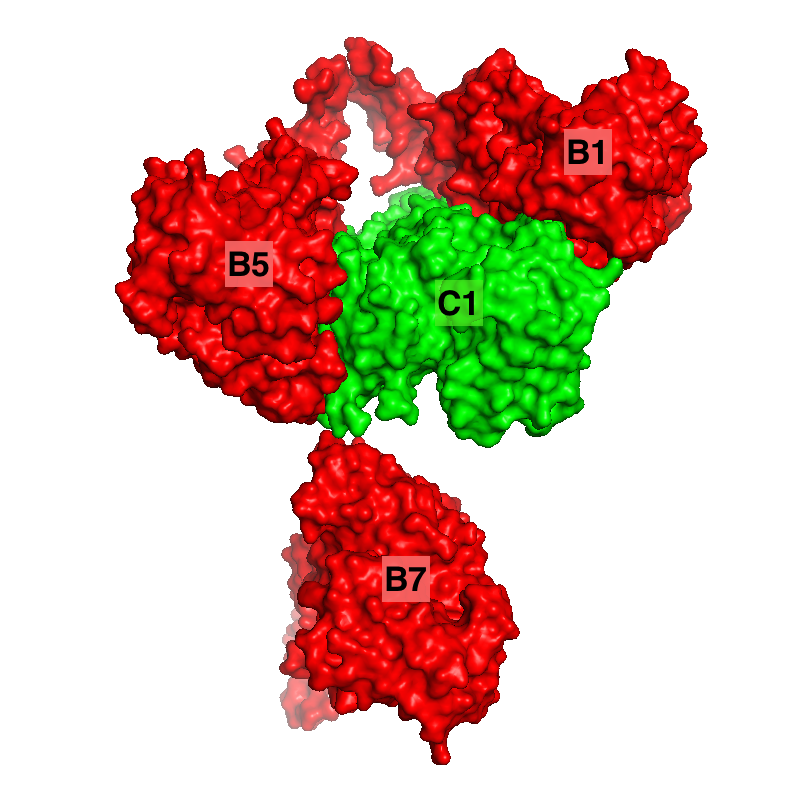}
 \includegraphics[width=0.3\linewidth]{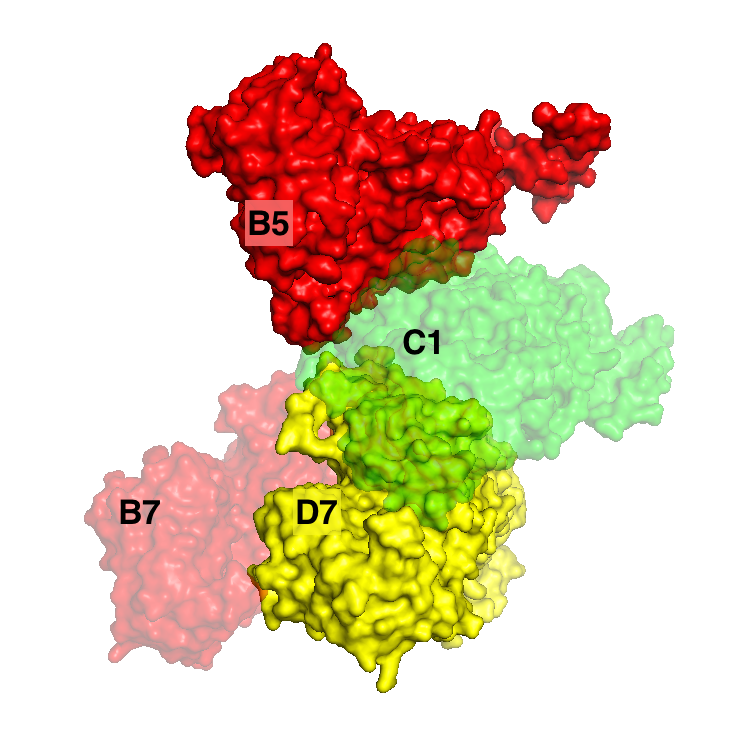}
 \includegraphics[width=0.3\linewidth]{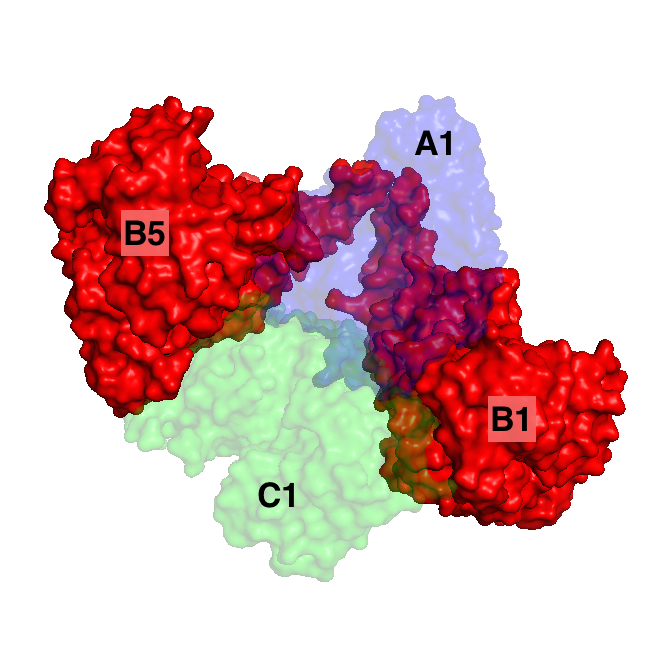}
 \caption{Labeled surface representation of B1-B5-B7-C1 (left), B5-D7 (center), and B1-B5 (right). Complexes have been rotated to show interfaces.}\label{fig:s-surfs}
\end{figure}

\begin{figure}
\includegraphics[width=0.48\linewidth]{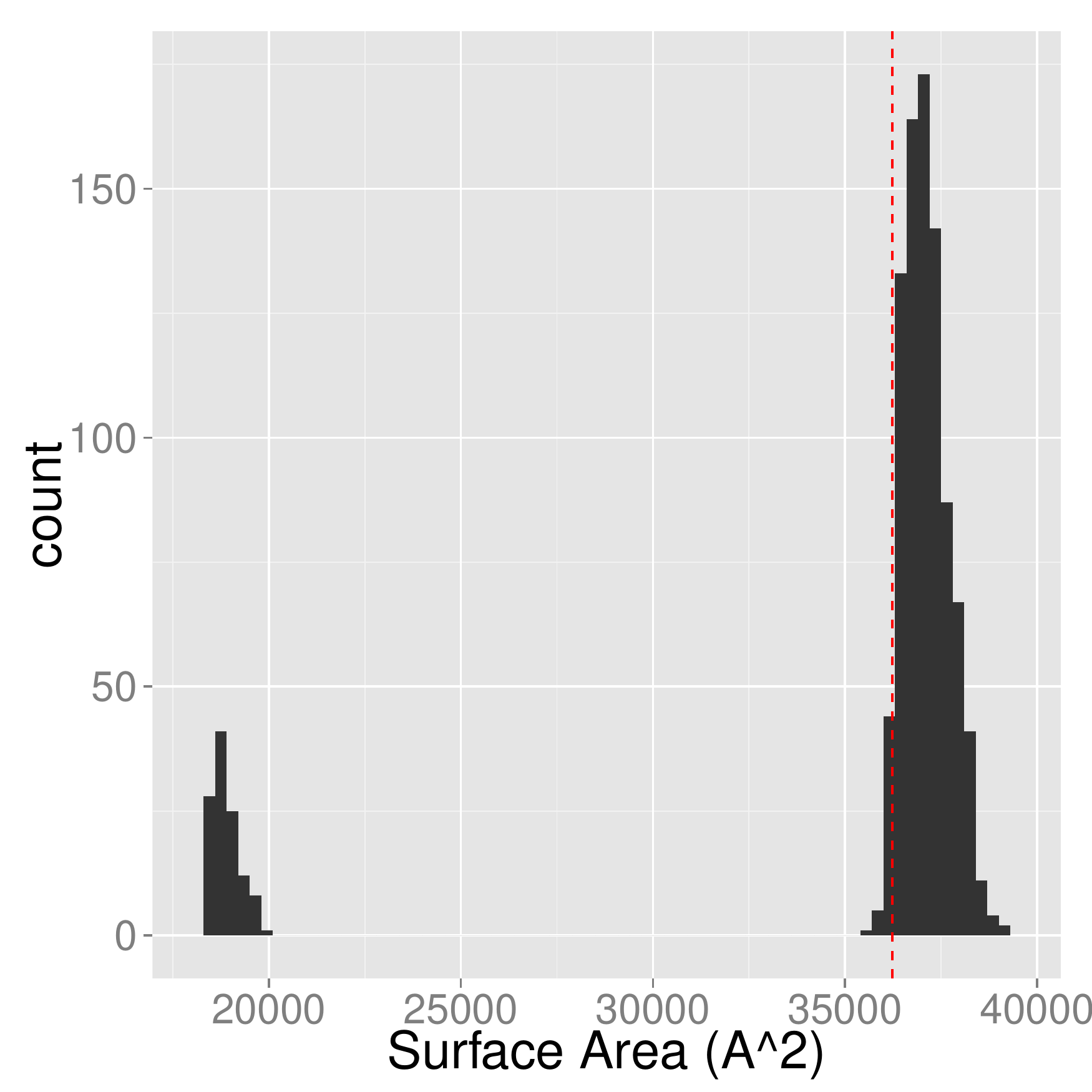}
\includegraphics[width=0.48\linewidth]{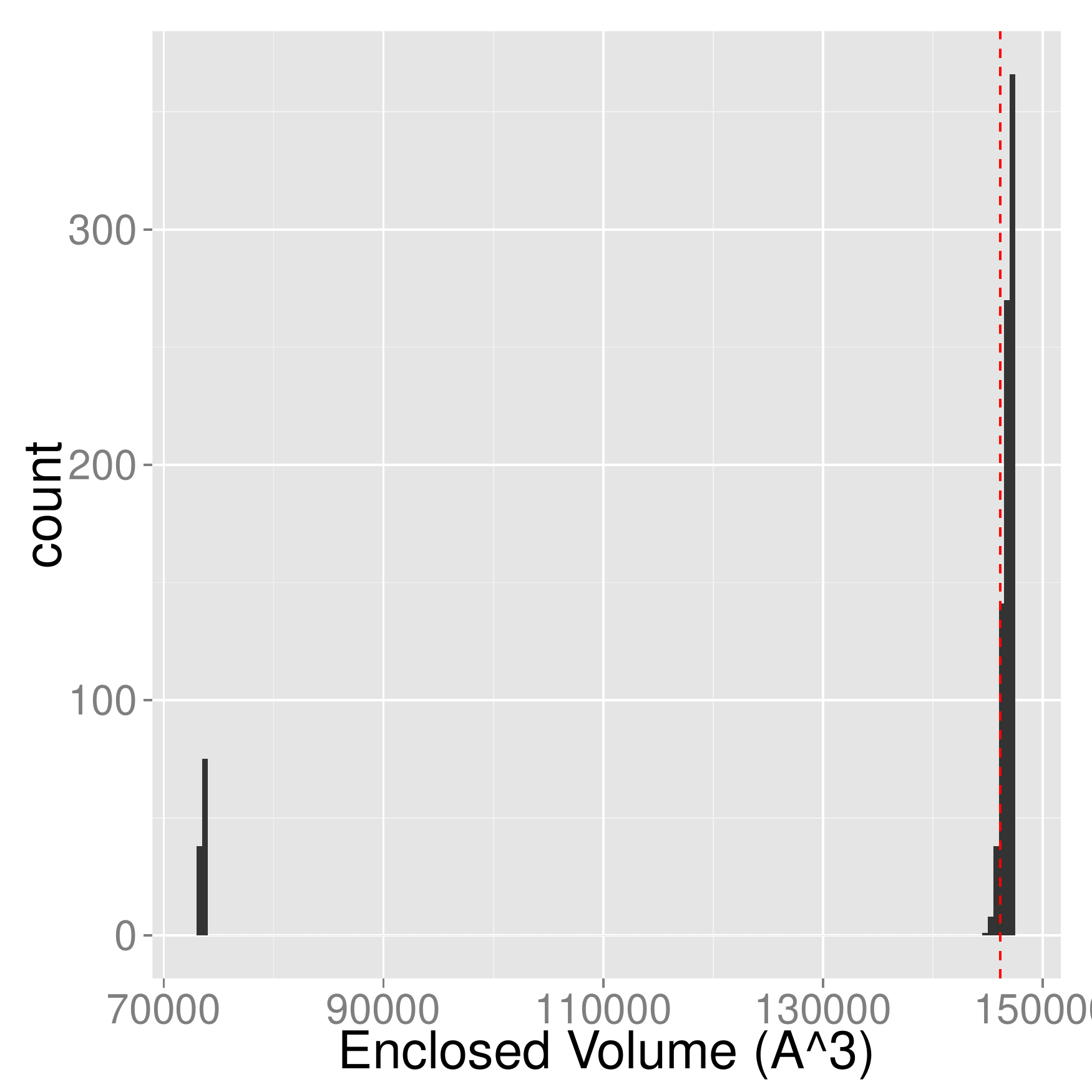}\\
\includegraphics[width=0.48\linewidth]{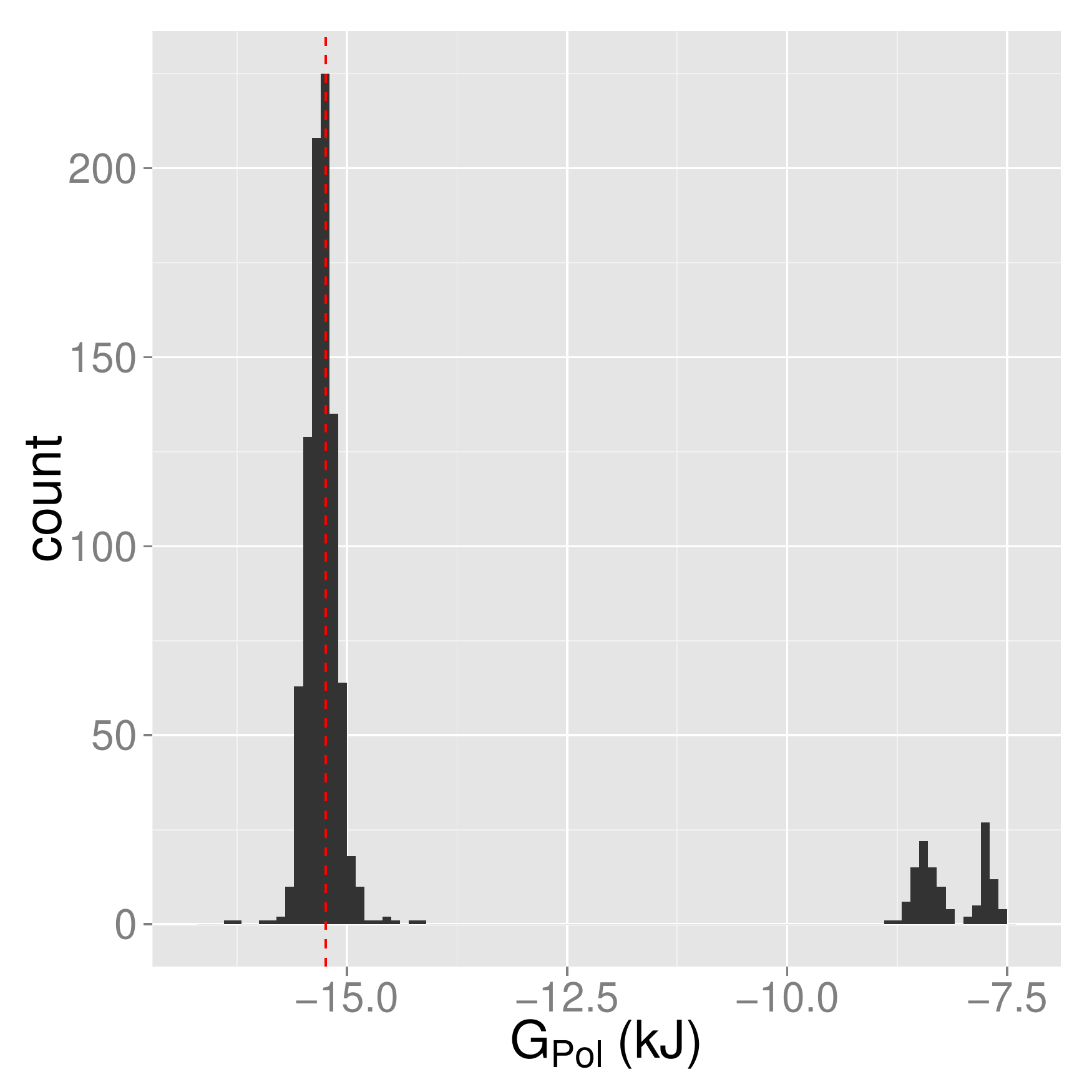}
\includegraphics[width=0.48\linewidth]{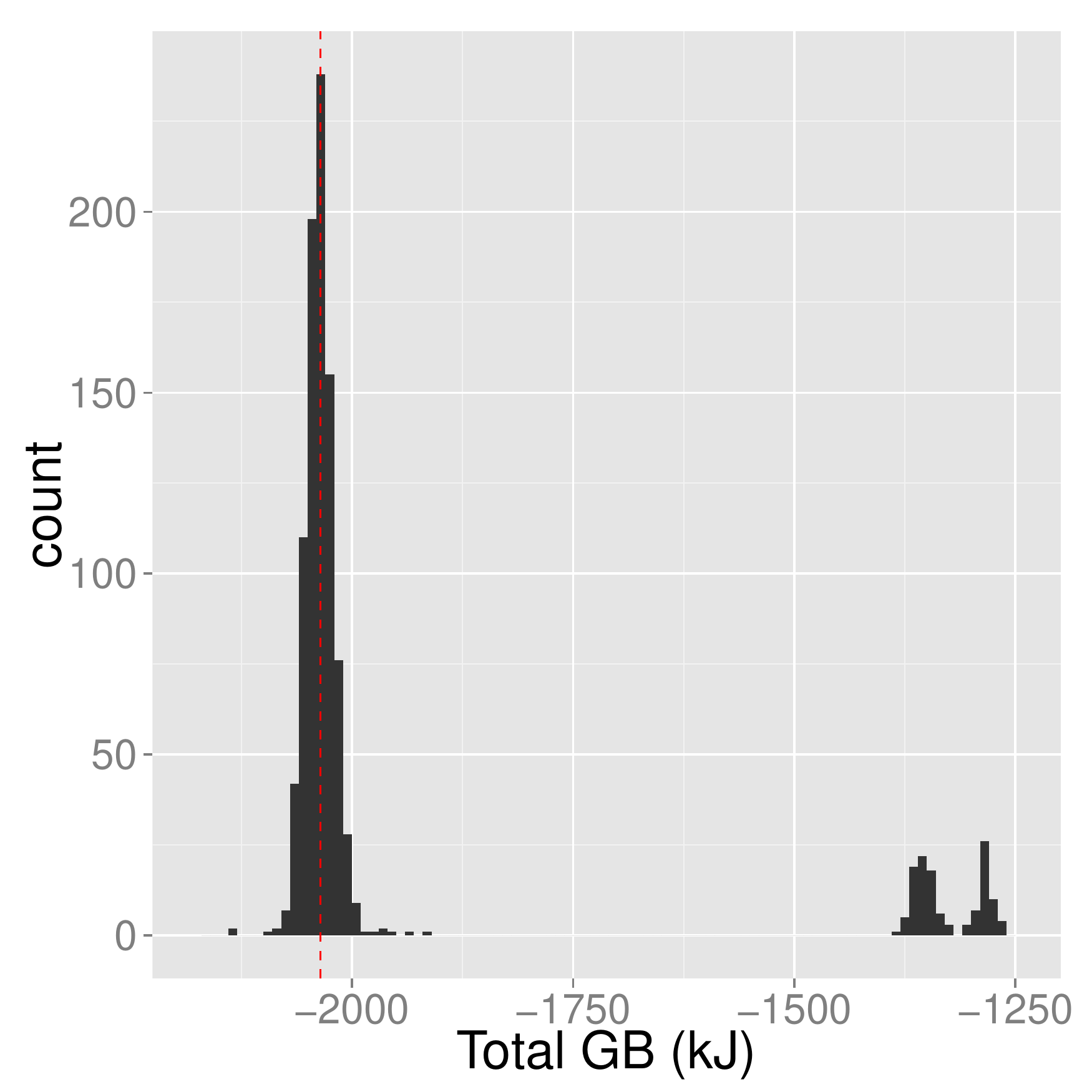}
\caption{Histogram plots of exposed surface area, enclosed volume, $G_{pol}$ and total energy (GBSA model) for all samples of B1-B5. Dotted vertical lines show the value computed on the un-sampled molecule. Bimodal nature of plots reveal low surface area between B1 and B5. See the surface representation in Figure \ref{fig:s-surfs}.}\label{fig:s-b1-b5}
\end{figure}

\begin{figure}
\includegraphics[width=\linewidth]{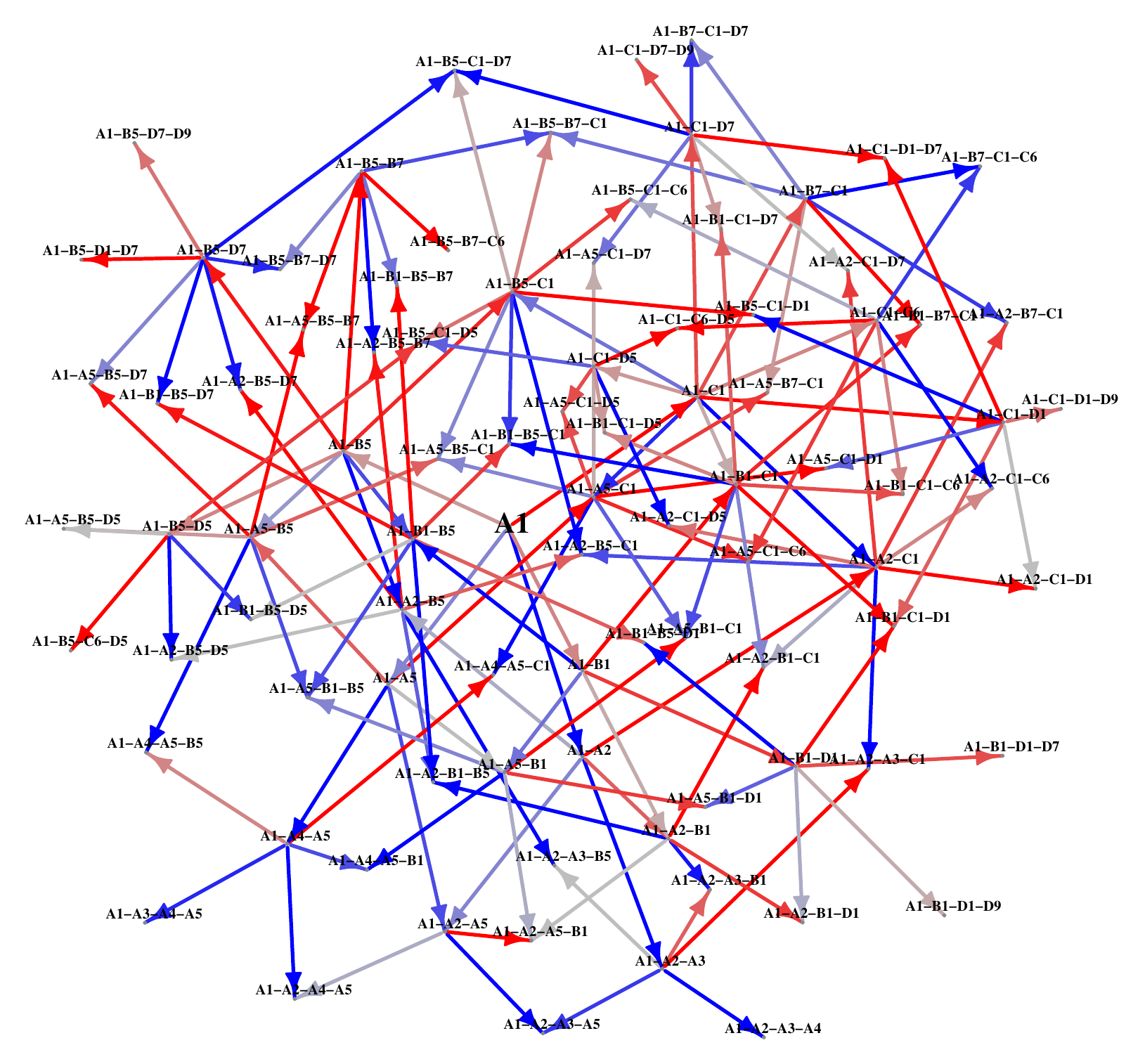}
\caption{Network graph of possible subassemblies formed, up to size 4, when starting from subunit A1. Color ranges from red (low Wilcoxon score) to grey to blue (high Wilcoxon score). Potential sinks are identified by nodes that have many incoming blue edges, and nodes with low probability of occurring have many incoming red edges.}
\end{figure}

\begin{figure}
\includegraphics[width=4in]{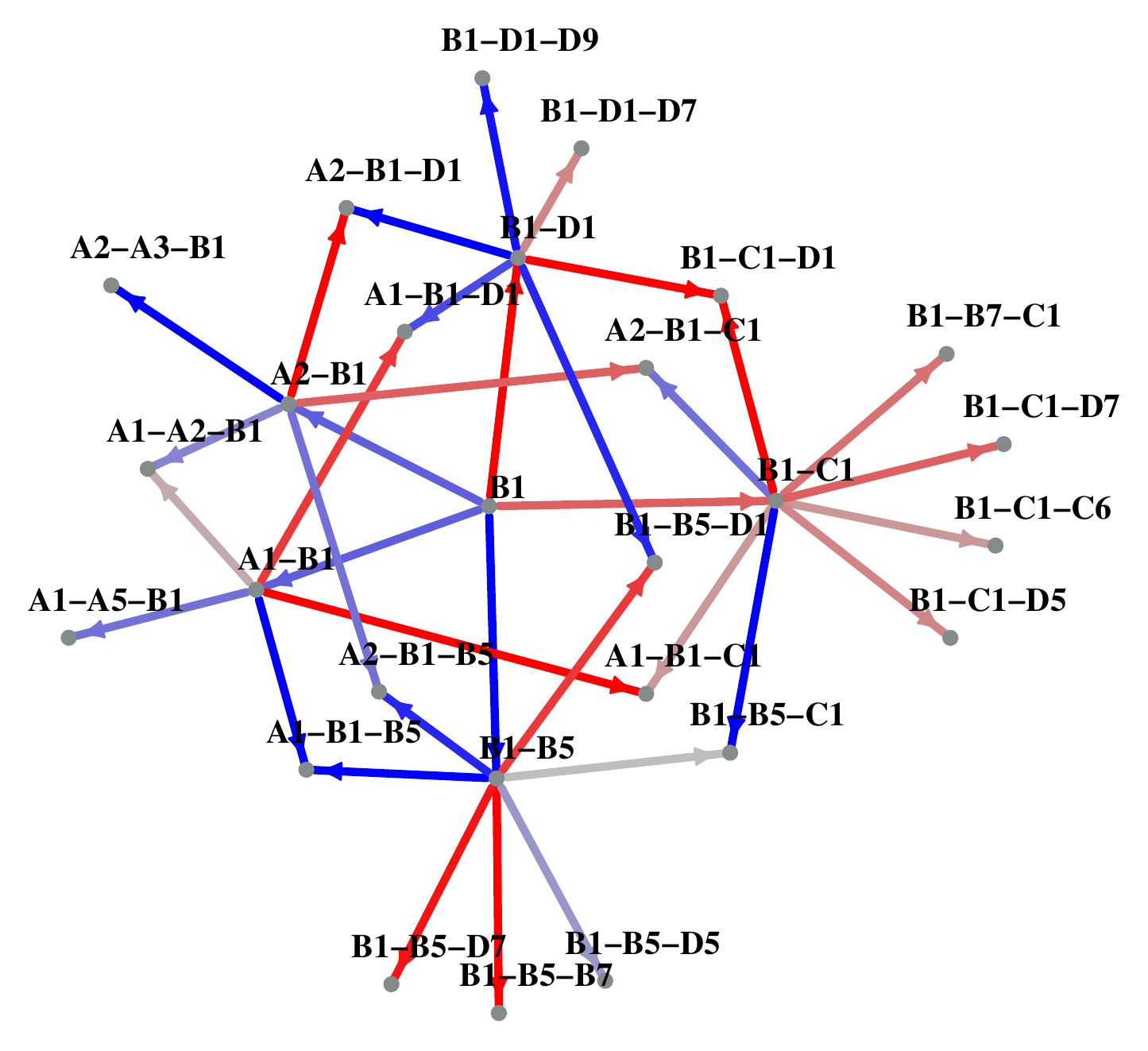}
\caption{Network graph of possible subassemblies formed, up to size 3, when starting from subunit B1. Color ranges from red (low Wilcoxon score) to grey to blue (high Wilcoxon score). Potential sinks are identified by nodes that have many incoming blue edges, and nodes with low probability of occurring have many incoming red edges.}
\end{figure}

\begin{figure}
\includegraphics[width=5in]{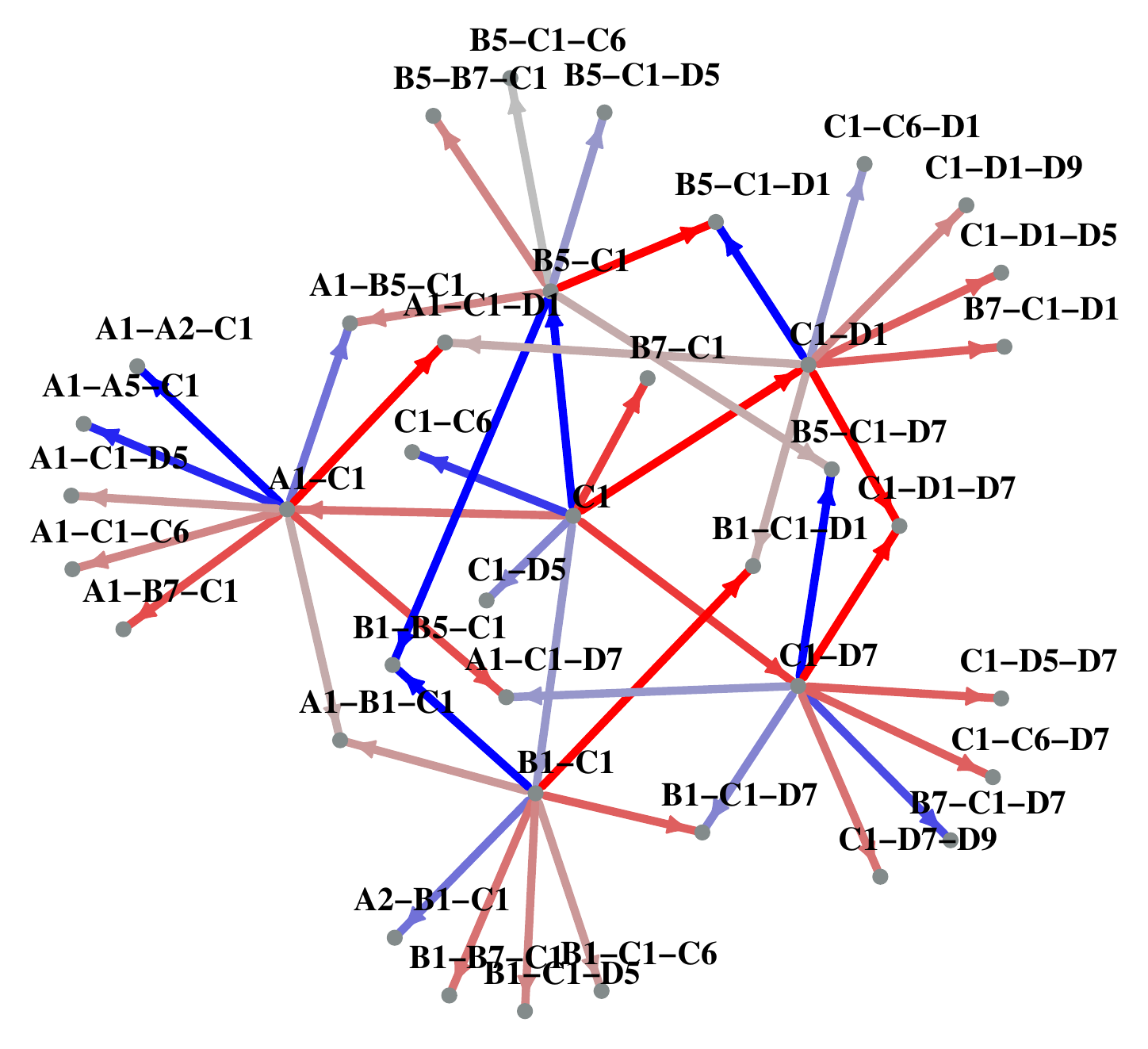}
\caption{Network graph of possible subassemblies formed, up to size 3, when starting from subunit C1. Color ranges from red (low Wilcoxon score) to grey to blue (high Wilcoxon score). Potential sinks are identified by nodes that have many incoming blue edges, and nodes with low probability of occurring have many incoming red edges.}
\end{figure}

\begin{figure}
\includegraphics[width=\linewidth]{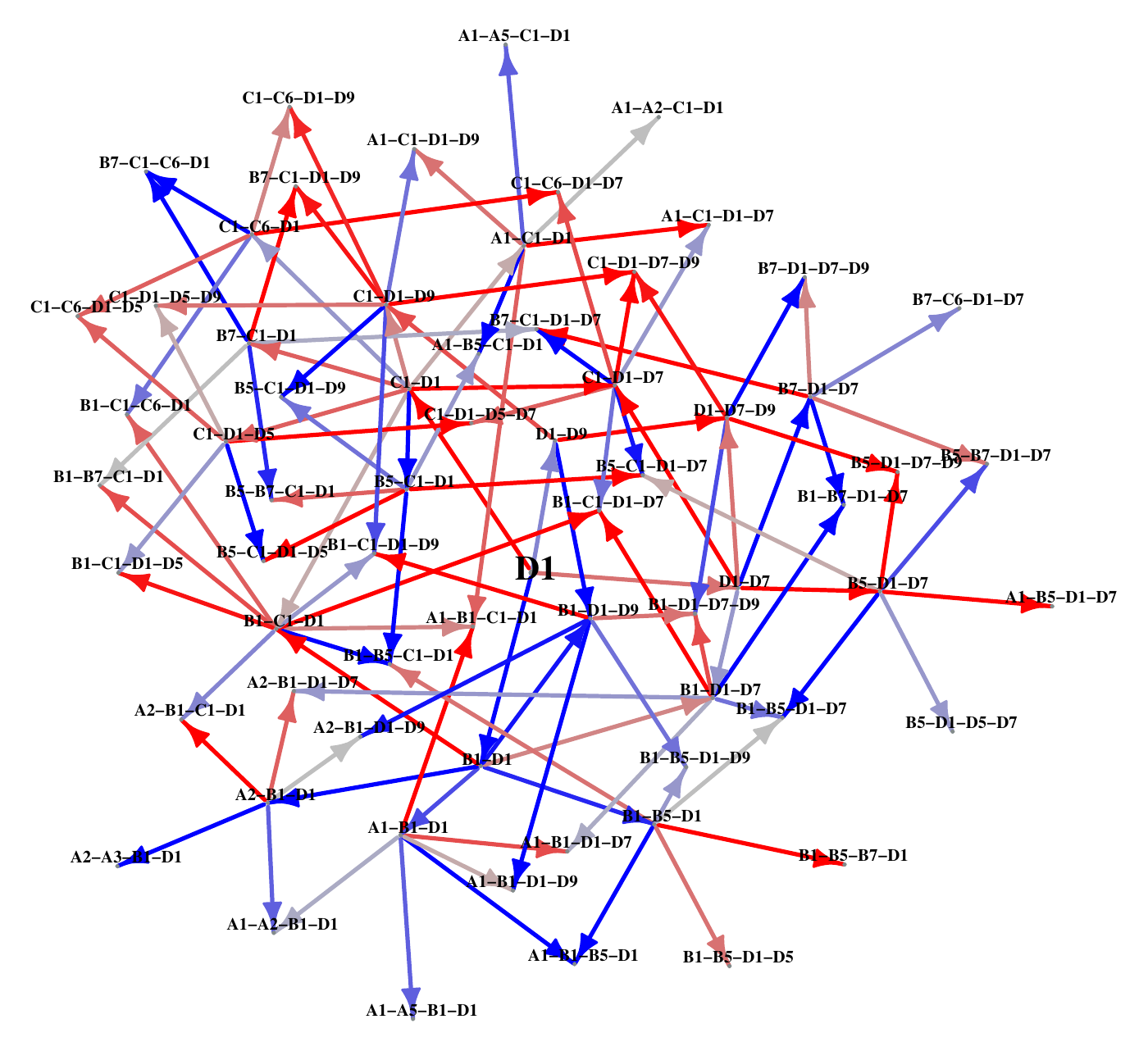}
\caption{Network graph of possible subassemblies formed, up to size 4, when starting from subunit D1. Color ranges from red (low Wilcoxon score) to grey to blue (high Wilcoxon score). Potential sinks are identified by nodes that have many incoming blue edges, and nodes with low probability of occurring have many incoming red edges.}
\end{figure}

%\bibliography{bib/virus,bib/others,bib/symmetry,bib/prior,bib/sampling,bib/uqrefs,bib/energy}
\bibliography{capsidUQ-arXiv}

\end{document}